\documentclass{aa}
\usepackage[varg]{txfonts}
\usepackage{graphicx}
\usepackage{amssymb}
\usepackage{lscape}
\newcommand{\HII}{\ion{H}{II}}

\newcommand{\CO}{$^{13}$CO}
\newcommand{\tCO}{$^{12}$CO}
\newcommand{\quot}[1]{
		\textquotedblleft #1\textquotedblright}

\bibpunct{(}{)}{;}{a}{}{,} % to follow the A&A style

\usepackage{natbib,twoopt}
\usepackage[breaklinks=true,draft]{hyperref} %% to avoid \citeads line fills
\makeatletter
\begin{document}

\title{Giant molecular filaments in the Milky Way}

\subtitle{Part two: The fourth Galactic quadrant}

\author{J. Abreu-Vicente\inst{1}\thanks{Member of the
International Max Planck Research School (IMPRS) at the University
of Heidelberg}$^{,}$\thanks{Table A.2 is only available in electronic form
at the CDS via anonymous ftp to cdsarc.u-strasbg.fr (130.79.128.5).}, S. Ragan\inst{2}, J. Kainulainen\inst{1}, 
Th. Henning\inst{1}, H. Beuther\inst{1}, K. Johnston\inst{2}.}

\authorrunning{Abreu-Vicente, J. et al.}
\titlerunning{Giant molecular filaments in the Milky Way: the fourth quadrant}

\offprints{J. Abreu-Vicente, \email{abreu@mpia-hd.mpg.de}}

\institute{$^{1}$Max-Planck-Institut f\"{u}r Astronomie (MPIA),
K\"{o}nigstuhl 17, 69117, Heidelberg, Germany\\
$^{2}$School of Physics and Astronomy, University of Leeds, Leeds, UK}
\date{\today}

%\abstract {} {Text of aims} {Text of methods} {Text of results} {}
%We also put   
%We extend the census of GMFs identified by NIR/MIR extinction features
%to the fourth Galactic Quadrant. We find that most of our GMFs 
%are not associated with spiral arms. This is in contradiction with 
%Li+2013, thus suggesting a selection effect in the GMFs identified with 
%both methods. We also analyze their physical properties, reporting a
%tentative correlation between the dense gas mass fraction of the
%GMFs and their position above the Galactic midplane.
\abstract{Filamentary structures are common morphological 
features of the cold, molecular interstellar medium (ISM). 
Recent studies have discovered massive, hundred-parsec-scale 
filaments that may be connected to the large-scale, Galactic 
spiral arm structure. Addressing the nature of these giant 
molecular filaments (GMFs) requires a census of their occurrence and properties.}
  % aims heading (mandatory)
   {We perform a systematic search of GMFs in the fourth 
   Galactic quadrant and determine their basic physical properties.}
  % methods heading (mandatory)
   {We identify GMFs based on their dust extinction signatures 
   in the near- and mid-infrared and the velocity structure probed by 
   $^{13}$CO line emission. We use the $^{13}$CO line emission 
   and ATLASGAL dust emission data to estimate the total and 
   dense gas masses of the GMFs. We combine our sample with an 
   earlier sample from literature and study the Galactic environment of the GMFs.}
  % results heading (mandatory)
   {We identify nine GMFs in the fourth Galactic quadrant. 
   Six in the Centaurus spiral arm and three in 
   inter-arm regions. Combining this sample with an earlier 
   study using the same identification criteria in the first 
   Galactic quadrant results in 16 GMFs, nine of which are 
   located within spiral arms. The GMFs have sizes of 80-160 
   pc and $^{13}$CO-derived masses between $5-90 \times 10^4$ M$_\odot$. 
   Their dense gas mass fractions are between 1.5-37$\%$, which is higher 
   in the GMFs connected to spiral arms. We also compare the different 
   GMF-identification methods and find that emission and extinction-based techniques overlap only partially, thereby highlighting the need to 
   use both to achieve a complete census.}
{}
\maketitle

\section{Introduction}\label{intro}

Filamentary structures are present in the interstellar 
medium (ISM) in a wide variety of environments over a wide range of scales, 
and they may play a major role in star formation~\citep[e.g., ][]{andrePPVI}.  
Especially the presence of parsec-sized filamentary
structures in the ISM and their close relationship with 
star formation have been known for several
decades~\citep[e.g., ][]{schneider79,bally87,beuther11,hacar13,kainul13,andrePPVI,molinariPPVI,stutz15}.  
These filaments are omnipresent in
molecular clouds, no matter whether these clouds are quiescent or
harbor star-forming activity \citep[e.g.,][]{molinariPPVI}.  Understanding the
physical origin and evolution of filaments is therefore needed to explain  the whole process of star formation.

The discovery of \quot{Nessie}~\citep{jackson10,goodman14}, an 80\,pc
long filament associated with the Scutum-Centaurus spiral arm, has initiated the
study of a family of giant molecular filaments (GMFs) in the Milky
Way. In addition to Nessie, four other GMFs had been discovered early on and 
studied~\citep{beuther11,kainul11,battersby12,li13,tackenberg13}.
All these five GMFs were identified first as absorption features against the 
mid-infrared (MIR) background of the Galaxy and confirmed to be continuous physical
objects using additional spectral line information. The existence of
such filaments raises the question whether they could be connected 
to the large-scale, Galactic spiral arm structure.
A systematic Galactic census of GMFs, characterizing
their occurrence and properties, is required to answer this question.

The first systematic study of GMFs was carried out by~\citet[][hereafter, R14]{ragan14}. 
Their study focused on the first Galactic quadrant. 
R14 identified a series of filamentary near-infrared (NIR) and MIR
extinction features. They used \CO ~spectral information to search 
for low-density gas bridges connecting the extinction features.
Their goal was to find the longest possible extent of the gas
connecting filamentary structures. They found seven GMFs 
with lengths between 50-230\,pc and masses on the order of
10$^{4}$-10$^{5}$\,M$_{\mathrm{\sun}}$. 
They used the Milky Way spiral-arm model of~\citet{vallee08}
to investigate the GMFs in the Galactic context. They
found that, unlike Nessie, six out of eight of their GMFs 
lie in inter-arm regions, rather than in spiral arms. 
 
\citet{goodman14} suggest that Nessie may be a part of much
longer structure that is located in the Galactic midplane
within a spiral arm. They call this structure a \quot{bone}.
Using a method similar to R14,~\citet{zucker15} report ten new
bones. These bones are smaller and less massive than the GMFs of R14.

Recently,~\citet{wang15} have approached the census of GMFs from a different 
perspective. They used dust emission from \textit{Herschel} to 
identify nine GMFs with masses and lengths similar to those found by R14.
Generally, both filament-finding methods do not  
identify the same filaments; if they do so, the size of 
the structures is not necessarily the same.
Using a more recent model of the spiral structure of
our Galaxy~\citep{reid14}, they found much higher coincidence between their GMFs 
and spiral arms than R14~\citep[who used the][model]{vallee08}, 
with seven out of nine filaments located in Scutum-Centaurus and Sagittarius 
spiral arms. 

Owing to the relatively low number of known GMFs and uncertainties 
in the Galactic models, the relation of GMFs to the Galactic structure 
remains an open question. Extending the census of GMFs to other
quadrants is key to obtain a Galaxy-wide piture of the
physical properties of the GMFs.

One key problem in identifying GMFs is that column density data 
(extinction or emission) alone are not sufficient; spectral line 
data are needed to ascertain that the structure has a continuous 
velocity pattern (and hence is likely a continuous object in three dimensions).
In short, the extinction patterns must be connected by a molecular gas tracer, 
usually \CO, and exhibit velocity coherence (i.e., the velocities
of the filaments must have no steep jumps, but rather show 
continuous velocity gradients, if any). 
The requirement of having \CO ~data greatly hampers building 
a systematic census of GMFs: an unbiased \CO ~survey exists 
only for the first Galactic 
quadrant~\citep[][Galactic Ring Survey, GRS hereafter]{jackson06}.
The GRS covered the region $17\degr\leq l \leq 55\degr$
and $b\leq|1\degr|$. The recent three-mm Ultimate Mopra Milky Way 
Survey~\citep[ThruMMS]{barnes-thrumms1,barnes-thrumms2}\footnote{http://www.astro.ufl.edu/~peterb/research/thrumms/} presents a
good opportunity to trace molecular cloud dynamics in 
the fourth Galactic quadrant.
This ongoing survey covers the fourth quadrant
in \tCO, ~\CO, ~C$^{18}$O, and CN.

In this paper, we extend the current census of GMFs to the fourth Galactic quadrant.  
We identify the GMFs as NIR/MIR extinction features that are connected structures 
in \CO ~data as probed by the ThruMMS survey. We present a sample of nine newly
identified GMFs and their physical properties. 
We place the results in the Galactic 
context with the help of models of the spiral-arm pattern of the Galaxy.
Finally, we compare the
different filament-finding methods to better understand their limitations and
complementarity with each others. 

\section{Data and methods}\label{sec:data}

\subsection{Data}

\subsubsection{\tCO ~and \CO ~data}\label{sec:totalGas}

We use the \CO (J$=$1--0) observations of the ThrUMMS survey 
DR3~\citep{barnes-thrumms1,barnes-thrumms2} 
to test the velocity coherence of the filament candidates.
We also use the data to estimate the distance to the
GMFs and to obtain their total masses.
This ongoing survey is observing the fourth Galactic quadrant
at latitudes $|b|<1\deg$ in \tCO, \CO, C$^{18}$O and CN with
an angular resolution of $72\arcsec$ and an approximate
rms of 1.5\,K\,km/s. ThruMMS offers full spectral coverage of the \CO ~line
at a spectral resolution $\sim0.3$\,km/s.
The observations of \tCO ~and \CO ~are mostly complete at $|b|<0.5\degr$.
However, less than 25\% is complete at Galactic latitudes $|b|>0.5\degr$.

\subsubsection{Dust continuum at $870\,\mathrm{\mu m}$ as a dense gas tracer}\label{sec:denseGasData}

We employ the ATLASGAL survey~\citep{schuller-09,csengeri14} to trace
the dense gas component of the GMFs.
This survey observed cold dust emission in a large area 
($-80\degr\leq l\leq 60\degr$) of the 
Galactic plane at $870\,\mu$m, with a FWHM
of 19.2$\,\arcsec$ and a $rms\sim50\,$mJy/beam. Although dust emission at sub-mm
wavelengths does not generally trace only
dense gas, ATLASGAL filters out large scale
(2.5$\arcmin$) emission, thus making the observations 
specially sensitive to the densest gas component, generally 
located in the cold interior of molecular clouds.

\subsubsection{Velocity data for the dense gas tracer}\label{sec:dense2}

Unfortunately, with only ATLASGAL data, we cannot 
know whether the emission arises from a GMF 
or from a different point along the line-of-sight.
We need extra spectral information. We employ several catalogs of star 
formation signposts to confirm that the dense gas is 
associated with the GMF. We search for counterparts of these
catalogs with ATLASGAL clumps and compare the velocity of
the sources with those of the GMFs.

We use the sources with radio recombination line counterparts
in the WISE catalog of \HII\,regions~\citep{anderson14},
the Red MSX Survey~\citep[RMS]{lumsden13}, and the catalogs of NH$_{3}$~\citep{purcell12}
and clumps H$_{2}$O masers~\citep{walsh11} of the HOPS survey.
In addition, we use a series of
follow-up studies of the ATLASGAL survey: the catalogs of CO depletion and
isotopic ratios~\citep{giannetti14}, methanol massers~\citep{urquhart13a}, 
and massive star-forming clumps~\citep{urquhart14b}.
Further, we use the the catalog of dense clumps from the MALT90 
survey~\citep{foster11,foster13,jackson13}. We list the dense gas
tracers associated to the GMFs in Table~\ref{tab:denseTracers}.

\subsection{Identifying giant molecular filaments}\label{sec:filCat}

We identify the filament candidates following the same procedure as in R14.
The first step is to identify filamentary extinction features by-eye
at MIR and NIR wavelengths in the Galactic plane (see Fig.~\ref{fig:apF3072_3054}--\ref{fig:apF3589_3574}). 
In this step, we use available data from GLIMPSE~\citep{benjamin03} and 
2MASS~\citep{Skrutskie06} surveys\footnote{Both surveys can be visualized 
in web interfaces at: \emph{http://www.alienearths.org/glimpse/} and 
\emph{http://aladin.u-strasbg.fr/AladinLite/}},
representing wide, unbiased and continuous coverage
of the Galactic plane at MIR and NIR wavelengths.

A group of 5 coauthors inspected the data by eye
searching for the extinction features.
The GMF candidates must satisfy two conditions:
\emph{1)} the extinction features must have a projected length
of $\gtrsim 1\degr$, and \emph{2)} the group members, in pairs of
two persons, must independently confirm the extinction feature as a filament-like
structure. Massive episodes of star-formation can disrupt filaments,
so we allow for gaps in the extinction structures if signs of 
massive star-formation are present (e.g., \HII ~regions).
We note that the photodissociation regions surrounding \HII ~regions
have strong PAH emission. As a result, NIR extinction features
may coincide with emission at MIR wavelengths (e.g, 8\,$\mu$m)
if there are \HII ~regions or strong radiation sources nearby~\citep{draine11}.
Following this procedure, we find 12 GMF candidates
within the fourth Galactic quadrant. The candidates
are listed in Table~\ref{tab:filCandList}.
In the next section, we explore whether the GMF candidates are 
physically connected using line emission data.

\begin{table}
\caption{Filament candidates in the fourth Galactic quadrant.} %
%\label{Table1} % is used to refer this Table in the text
\centering % used for centering Table
\begin{tabular}{l c c c c} % centered columns (4 columns)
\hline\hline % inserts double horizontal lines
Candidate ID	&	$l_{\mathrm{ini}} [\degr]$	&	$l_{\mathrm{end}} [\degr]$	&	$b_{\mathrm{ini}} [\degr]$	&	$b_{\mathrm{end}} [\degr]$	\\\hline
F358.9-357.4	&	358.9	&	357.4	&	-0.4	&	-0.4	\\
F354.7-349.7\tablefootmark{a}	&	354.7	&	349.7	&	0.4	&	0.5	\\
F343.2-341.7	&	343.2	&	341.7	&	0.0	&	0.4	\\
F341.9-337.1	&	341.9	&	337.1	&	-0.2	&	-0.4	\\
F335.6-333.6	&	335.6	&	333.6	&	-0.2	&	0.4	\\
F329.3-326.5	\tablefootmark{a}&	329.3	&	326.5	&	-0.3	&	0.0	\\
F329.4-327.1	\tablefootmark{b}&	329.4	&	327.1	&	-0.3	&	1.4	\\
F326.7-325.8	\tablefootmark{b}&	326.7	&	325.8	&	0.9	&	-0.2	\\
F324.5-321.4	&	324.5	&	321.4	&	-0.5	&	0.1	\\
F319.0-318.7	&	319.0	&	318.7	&	-0.1	&	-0.8	\\
F309.5-308.7	&	309.5	&	308.7	&	-0.7	&	0.6	\\
F307.2-305.4	&	307.2	&	305.4	&	-0.3	&	0.8	\\
\label{tab:filCandList}
\end{tabular}
\tablefoot{
\tablefoottext{a}{These filament candidates have not been confirmed
		as GMFs because they are not velocity coherent.}
\tablefoottext{b}{There is only ThruMMS coverage for the part of the filament.}
		}
\end{table}

\subsubsection{Velocity coherence of the candidates}\label{sec:veloExt}

To avoid projection effects caused by physically
unrelated molecular clouds along the line-of-sight,
we only consider the candidates listed in Table~\ref{tab:filCandList} 
as GMFs once it is confirmed that they are
velocity coherent structures.
Very long filaments are likely to show velocity
gradients due to the differential rotation of the Galaxy. 
These gradients depend on the location of the filament in the Galaxy. 
For this reason, we do not restrict the velocity range of the filament candidates,
but rather require them to show continuous velocity variations, without
steep jumps. 
The filaments satisfying this requirement are considered velocity coherent.

We test the velocity coherence of the candidates using \CO ~observations 
of the ThrUMMS survey (see Sect.~\ref{sec:totalGas}). 
We create position-velocity (PV) diagrams for each GMF candidate
integrating the full spectral coverage collapsed over the 
whole latitude axis. In this step, we used the 
function \texttt{sum} of \texttt{python}.
In the cases in which the PV diagrams show a series of different PV-components,
we create \CO ~integrated intensity maps for each component.
We then compare the extinction features 
with the \CO ~integrated intensity maps. 
If they represent a single \CO ~structure,
then the filament candidates are labeled GMFs. 

The following candidates are eliminated because their extinction features have 
no single coherent \CO ~velocity component: F354.7-349.7 and F329.3-326.5.
The candidates F329.4-327.1 and F326.7-325.8
could not be confirmed because of the lack of proper coverage in the ThrUMMS data.
The remaining nine candidates were classified as GMFs.

The spiral arms are seen as seen as a single \CO ~component
with a wide-velocity range
in the PV diagrams (see Fig.~\ref{fig:apF3072_3054}--\ref{fig:apF3589_3574}).
If our GMFs lie inside spiral arms, then it is
possible that we are integrating over the full spiral arm,
thus overestimating the velocity range of the GMF. 
We do a fine tuning of the velocity range
of the GMFs. We created position-velocity (PV) diagrams over a line 
following the identified extinction features for each candidate.
This process can be done with the \texttt{python} tool,
\texttt{Glue}\footnote{http://www.glueviz.org/}.
These PV diagrams are shown in the bottom panels of
Fig.~\ref{fig:apF3072_3054}--\ref{fig:apF3589_3574}. 
We also show in Table~\ref{tab:lbvtracks} the l--b--v
tracks of each filament.

\subsubsection{Biases in the extinction-based filament finding method}\label{sec:biasExt}

We acknowledge that this method is necessarily subjective. 
We make an effort to reduce the subjectivity by requiring that at least
three group members agree with the filament identification. 
In addition, some GMFs that could have been potentially identified
may have been missed by our search approach. 

Our filament finding method is biased 
towards the identification of quiescent structures (R14). 
Even though we allow for gaps in filaments if they carry signs
of massive star formation, such violent episodes can disrupt
molecular clouds, making them difficult (if not impossible) 
to be identified as GMFs. 

The observation of extinction features against NIR and MIR 
background requires intense background emission to 
have enough contrast to identify extinction features. This is true at low Galactic
latitudes, where the star density is high, but it is not the case at 
high latitudes. In general, this is not a severe issue in this work since
we target specifically the Galactic plane. However, the region $325\degr<l<320\degr$
shows low background emission at MIR wavelengths. The 
identification of extinction features in this region is therefore more
difficult due to the lack of contrast between the background
and the dense foreground structures. 

\subsection{Estimating the total and dense molecular gas masses}

\subsubsection{Total molecular gas mass}\label{sec:totalMass}

To calculate the total gas mass of the filaments, we first 
obtain the column densities of \CO. We do so following 
a standard scheme in which the kinetic temperature of the \CO ,~$T_{^{13}\mathrm{CO}}$, is 
assumed to be the same as that of \tCO ,
and equal to the kinetic temperature, $T_{\mathrm{k}}$~\citep[e.g.,][]{tools}.
We obtain $T_{\mathrm{k}}$ from the brightness temperature of the optically thick
line \tCO , $T_{\mathrm{B,^{12}CO}}$,

\begin{equation}
T_{\mathrm{k}} = \frac{5.5}{\mathrm{ln}(1+\frac{5.5}{T_{\mathrm{B,^{12}CO}}+0.82})},
\end{equation}

\noindent with $T_{\mathrm{B,^{12}CO}}$ obtained from the peak of the 
\tCO ~emission. The optical depth of the \CO ~line,
$\tau_{\mathrm{^{13}CO}}$, is obtained via

\begin{equation}
\tau_{\mathrm{^{13}CO}} = -\mathrm{ln} \left [   1 - \frac{T_{^{13}\mathrm{CO}}/5.3}{(e^{\frac{5.3}{T_{\mathrm{k}}}}-1)^{-1} - 0.16} \right ].
\end{equation}

We integrate the \CO ~spectra with peaks over 1.5\,K\footnote{
This value represents the rms of the \CO ~spectra.} 
to obtain the total column density of \CO 

\begin{equation}
N_{\mathrm{^{13}CO}} = 3.0\times10^{14}\frac{T_{\mathrm{k}}e^{\frac{5.3}{T_{\mathrm{k}}}}\int \tau_{\mathrm{^{13}CO}}(\nu)\mathrm{d}\nu}{1-e^{\frac{-5.3}{T_{\mathrm{k}}}}}.
\end{equation}

For consistency with R14, we convert $N_{\mathrm{^{13}CO}}$ into \tCO 
~column densities, $N_{\mathrm{^{12}CO}}$, using a $^{12}$CO/$^{13}$CO 
ratio that varies with galactocentric distance following the linear relation 
$^{12}\mathrm{CO}/^{13}\mathrm{CO}=5.41R_{\mathrm{gal}}[\mathrm{kpc}]+19.3$~\citep{milam05}. The galactocentric distances of the filaments 
are listed in Table.~\ref{tab:filCandVelo}.
Finally, the \tCO ~column densities are converted into column densities 
of molecular gas following: $N_{\mathrm{^{12}CO}}$/$N(\mathrm{H}_{2})=1.1\times10^{-4}$
~\citep{pineda10}. We find that \CO ~integrated intensities
of 1.5~K / km/s correspond to $N(\mathrm{H}_{2})\sim1.3\times10^{21}\mathrm{\,cm^{-2}}$.

Finally, we obtain the total molecular gas mass of the GMFs via
\begin{equation}
M_{\mathrm{total}} = N_{\mathrm{^{13}CO}} m_{\mathrm{H_2}} N_{\mathrm{pix}} A_{\mathrm{pix}} d^{2}, 
\end{equation}

\noindent where $m_{\mathrm{H_2}}$ is the mass of the molecular hydrogen,
$N_{\mathrm{pix}}$ and $A_{\mathrm{pix}}$ the number and area
of the pixels inside the GMF respectively, and $d$ the distance
to the GMF (See Sect.~\ref{sec:dist}).

\subsubsection{Dense gas mass}\label{sec:denseGas}

We used the ATLASGAL data to estimate the dense gas mass
of the GMFs. We required the ATLASGAL emission to be detected 
at 5$\sigma$ (250\,mJy/beam) for consistency with R14. This emission is equivalent to 
$N(\mathrm{H}_{2})=7\times10^{21}\mathrm{\,cm^{-2}}$.
We estimate the dense gas mass of each GMF via
\begin{equation}\label{mass}
M_{\mathrm{dense}} = \frac{Rd^{2}F_{870\mathrm{\mu m}}}{B_{870\mathrm{\mu m}}(T_{dust})\kappa},
\end{equation}

\noindent  where $F_{870\mathrm{\mu m}}$ is the ATLASGAL flux, 
$d$ is the distance to the filament,
and $B_{870\mathrm{\mu m}}(T_{\mathrm{dust}})$ is the blackbody radiation at 
870\,$\mu$m as a function of temperature, $T_{\mathrm{dust}}$, 
which we assume $T_{\mathrm{dust}}=20$\,K. 
$R=150$ is the gas-to-dust ratio~\citep{draine11}.
We used a dust absorption coefficient
$\kappa=1.42\mathrm{\,cm^{2}g^{-1}}$ at $870\,\mathrm{\mu m}$,
extrapolated from the dust model of~\citet{oh94} for dust
grains with thin ice mantles and a mean density of $n\sim10^{5}$\,cm$^{-3}$. 

It is possible that the ATLASGAL emission along the
line-of-sight of a filament is not related to the GMF, but 
rather with molecular clouds at different distances. 
To avoid this line-of-sight confusion we use spectral information
of the star-forming signposts introduced in Sect.~\ref{sec:dense2}.
If any of these signposts is associated to an ATLASGAL clump 
in projection and its $v_{\mathrm{LSR}}$\footnote{LSR stands for
local standard of rest.} lies inside the 
velocity range of the GMF, then we assume that this ATLASGAL
clump is part of the GMF. If we found no star-forming signposts
in an ATLASGAL clump, we also assume it to be part of the GMF.
If the $v_{\mathrm{LSR}}$ of the signposts associated with ATLASGAL emission 
is outside the range of velocities of the GMF, then the associated 
ATLASGAL clumps are neglected.

\section{Results}\label{results}

\subsection{Physical properties of the GMFs}

In this section we present physical properties of the 
GMFs: length, velocity gradient, total mass, dense gas mass, and
dense gas mass fraction. 

\subsubsection{Kinematic distance, length and velocity gradient}\label{sec:dist}

Before obtaining the mass and length of the GMFs, we 
estimate their kinematic distances using \CO ~data.
We fit the spectrum of each GMF with a Gaussian
and define the $v_{\mathrm{LSR}}$ as the centroid of the fit.
We used the model of the Galactic spiral arm pattern by~\citet{reid14}
to convert $v_{\mathrm{LSR}}$ into kinematic distances,
assuming the standard Galactic parameters. 
Our filament-finding method favors identification of 
nearby filaments. Although infrared dark clouds (IRDCs) can be seen
against the MIR background at both sides of the Galaxy,
only those in the near side would appear as clear extinction features
at NIR wavelengths~\citep{kainul11}. Our extinction method is therefore limited
to find filaments up to $\sim$8\,kpc distance~\citep{kainul11}.
We therefore assumed the near kinematic distances to our GMFs.
The velocities and distances to the GMFs are listed in Table~\ref{tab:filCandVelo}.
We find distances between 2.2--3.7\,kpc. 
These are similar to previously identified GMFs.
For comparison with the kinematic distances, we also list
in Table~\ref{tab:filCandVelo} the distances derived from dust 
extinction~\citep{marshall09}. We estimated the mean extinction distance
to the GMFs as the average of every counterpart in~\citet{marshall09}
associated to the GMFs and assume their standard deviation as the 
uncertainty. In general we obtain larger distances using this method.
This is consistent with a systematic offset of 1.5\,kpc between
kinematic and extinction derived distances, already reported
in~\citet{marshall09}.

We estimate the angular length of the GMFs using a 
line that follows the extinction and emission features at 
$8\,\mathrm{\mu m}$ from end to end of the significant \CO ~emission 
($\geq$1.5\,K\,km/s) of each filament 
(see Fig.~\ref{fig:apF3072_3054}--\ref{fig:apF3589_3574}). The significant
emission is estimated measuring the noise of the CO
integrated intensity maps.
We find angular lengths between $1\degr$ and $3\degr$.
The angular length is converted into physical length using the
distances previously estimated. No corrections are applied for the 
projection effects. These lengths are therefore lower limits. 
We found GMF projected lengths between 40--170\,pc, 
with a mean of $\sim$100\,pc. These values 
are similar to the filaments identified by R14.

We estimate the projected velocity gradient of the 
GMFs as  $\nabla v=(v_{\mathrm{ini}}$-$v_{\mathrm{end}})/l$,
where $v_{\mathrm{ini}}$ and $v_{\mathrm{end}}$ are the velocity 
centroids at both ends of the GMF and $l$ is the projected length of the GMF.
Most of the GMFs exhibit projected velocity gradients throughout their 
extent, except the GMF 324.5-321.4, that shows no projected velocity gradient 
over its 120\,pc size. We found projected velocity gradients between 
0--120\,km/s\,kpc$^{-1}$ (see Table~\ref{tab:filCandVelo}).
We emphasize that these velocity gradients are projected. 
We did not correct them from projection effects.
Therefore, these gradients offer a pure
observational measure and should not be directly connected
to velocity gradients introduced by Galactic rotation or shear motions.

\begin{table*}
\caption{Confirmed GMFs in the fourth Galactic quadrant.} %
%\label{Table1} % is used to refer this Table in the text
\centering % used for centering Table
\begin{tabular}{c c c c c c c c c c c} % centered columns (4 columns)
\hline\hline % inserts double horizontal lines

Filament ID	& $l$  & $b$   & $\theta$           &$v_{\mathrm{ini}},v_{\mathrm{LSR}},v_{\mathrm{end}}$            & $d_{kin}$\tablefootmark{d}     & $d_{ext}$  	                & $L$ 	&$\langle\nabla v\rangle$ 	& R$_{\mathrm{gal}}$ & $\langle z\rangle$ \\
      GMF          &[$\degr$]      &[$\degr$]       &[$\degr$]             &[km\,s$^{-1}$] & [kpc] & [kpc] &[pc]           &[km\,s$^{-1}$\,kpc$^{-1}$]      & [pc] & [pc]\\\hline
307.2-305.4	&[307.4,304.7]  &[-0.3,0.6]      &3.1            &-29, -35, -39	 &3.1$\pm0.7$&$5.7\pm0.9$	&168	 &59 & 7.2 & 20	                 \\
309.5-308.7	&[309.6,308.6]	&[-0.7,0.7]      &1.5            &-41, -43, -47	 &3.7$\pm0.7$&$4.7\pm0.3$   &97 	&62	& 6.8 & 25	              \\
319.0-318.7&[319.0,317.3]	&[-0.8,0.2]      &2.6            &-35, -40, -44	 &2.8$\pm0.4$&$4.9\pm0.4$   &73	&123	 & 6.7 & 25                \\
324.5-321.4	&[324.3,321.4]   &[-0.3,0.2]&3.1        &-32, -32, -32	 &2.2$\pm0.4$&---\tablefootmark{c}	   &119	&0 & 6.9 & 7      \\
324.5-321.4b\tablefootmark{a}	&[322.7,321.4]   &[-0.2,0.2]&1.0 &-32, -32, -32	 &2.2$\pm0.4$&$2.9\pm0.3$\tablefootmark{c}	           &38	&0 & 6.9 & 21		\\ 
335.6-333.6a	&[335.8,332.2]	&[0.0,-0.6]      &2.0      &-46, -50, -53	 &3.4$\pm0.3$&$4.5\pm0.5$                	&119	 &59	 & 5.7 & 7	                 \\
335.6-333.6b	&[332.8,331.6]	&[-0.2,-0.2]     &1.4      &-48, -50, -51	 &3.3$\pm0.3$&$5.8\pm0.5$                 	&81	&39	&5.8 & 13                 \\
341.9-337.1	&[342.2,340.2]	&[-0.1,-0.2]     &2.1       &-43, -44, -48	 &3.5$\pm0.3$&$5.1\pm0.5$                 	&128 	&31	 & 5.3 & 7                 \\
343.2-341.7	&[342.8,341.7]	&[0.0,0.5]       &1.5        &-43, -41, -41       &3.4$\pm0.3$&$4.9\pm0.3$                	&89	&22	& 5.3 & 33	                 \\ 
358.9-357.4	&[358.1,357.4]  &[-0.4,-0.3]     &1.6       &5, 7, 8	         &---\tablefootmark{b}&$2.9\pm0.3$        &81       &37	  & 5.6 & 3               \\
\label{tab:filCandVelo}
\end{tabular}
\tablefoot{Column 1: GMF identification; Cols. 2-3: Galactic latitude and longitude ranges of the GMFs;
  Col. 4: Angular length of the filament from end to end; Col. 5: $v_{\mathrm{ini}}$ and $v_{\mathrm{end}}$ represent the velocity centroid of the \CO ~emission
  at the ends of the filament.
  $v_{\mathrm{LSR}}$ is the velocity centroid of the \CO ~emission along the entire GMF. This value is used to estimate the kinematic distance to the GMF; Col. 6: Kinematic distances and uncertainties, obtained from the Galactic model of~\citet{reid14}; Col. 7:
  Distances obtained from dust extinction models~\citep{marshall09}. The distances were obtained as the mean of all the counterparts in the catalog~\citep{marshall09}, overlapping the GMFs. The uncertainties show the standard deviation.; Col. 8: Projected length of the filaments; Col. 9: Projected velocity gradient of the filaments, obtained as the ratio $(v_{\mathrm{end}}-v_{\mathrm{ini}})$/length;
  Col. 10: Galactrocentric radius; Col. 11: Mean height above the Galactic midplane in the middle of the filament.\\
  \tablefoottext{a}{Most conservative definition of GMF324.5-321.4 (See Sect.~\ref{sec:F323}).}
  \tablefoottext{b}{We avoid the use of kinematic distance in this GMF since it lies in the Galactic center region.}
  \tablefoottext{c}{No~\citet{marshall09} counterparts in this filament.}
  \tablefoottext{d}{We have used the kinematic distances to the sources, with the exception of
  					GMF 358.9-357.4, for which we have used the extinction-based distance estimate.}
}
\end{table*}

\subsubsection{Dense gas mass fraction (DGMF)}\label{sec:result-mass}

We estimate the total molecular gas mass and the dense gas
mass of the GMFs following the procedures described
in Sect.~\ref{sec:totalMass} and Sect.~\ref{sec:denseGas}.  
We obtain total molecular cloud masses 
in the range of $[1.4,9.4]\times10^{5}$\,M$_{\sun}$.
We found dense gas masses between $[2.1,310]\times10^{3}$\,M$_{\sun}$. 

We calculate the dense gas mass fraction (DGMF) of the GMFs as
the ratio of the dense to total molecular gas masses of the GMFs.
This quantity has been recently connected to the 
star-forming rate of molecular clouds~\citep{heiderman10,lada10,lada12,abreu15}.
In general, we find DGMFs between 1.5\% and 15\%. However,
the filament GMF 335.6-333.6 shows a larger DGMF, 37\%.
This large value is related to the massive \HII ~complex
G333, from which most of the ATLASGAL emission of the GMF
arises (see Fig~\ref{fig:ap3356_3336}). The DGMF of this filament
agrees with values found in massive ($\geq10^{5}$\,M$_{\sun}$) \HII ~regions
in the Galactic plane~\citep{abreu15}.
The lower DGMF values found in the other filaments are consistent
with those found in the GMF sample of R14 and other large
molecular filaments~\citep{battersby12,kainul13}. These values are also 
characteristic for local star-forming clouds~\citep{kainul09,lada10,lada12}
and high-mass star-forming clumps~\citep{johnston09,battisti14}.

The main uncertainty in our dense gas mass estimates
is the distance to the filaments. This is also the case 
for the total gas mass. Unfortunately, the ThruMMS data has a 
non-uniform noise coverage that may lead to lose significant 
\CO ~emission in some of our GMFs, limiting our ability to measure 
their total gas mass. In addition, five of the GMFs 
are only partly covered by the ThruMMS data. 
As a consequence of these issues, we can 
only measure lower limits of the actual total gas mass 
in most of our GMFs, resulting in upper limits for the DGMFs. 
These are indicated in Table~\ref{tab:filCandVelo2}. 
We also show these regions in Fig.~\ref{fig:apF3072_3054}--\ref{fig:apF3589_3574}. 

\begin{table*}
\caption{Masses and associations of the GMFs} %
%\label{Table1} % is used to refer this Table in the text
\centering % used for centering Table
\begin{tabular}{c c c c c c} % centered columns (4 columns)
\hline\hline % inserts double horizontal lines

Filament ID	&	M($^{13}$CO) 	&	M(AGAL)  &	DGMF     & Assoc.& Arm\\
	&	[$10^{4}$\,M$_{\sun}$]	&	 [$10^{4}$\,M$_{\sun}$] &	[\%]    & &\\\hline
GMF 307.2-305.4	&	$82$			&$8.7$				&	10.6	     &  G\,305 \HII ~complex     & Centaurus\\
GMF 309.5-308.7	&	$>81$		&$1.7$				&	$<2.1$      & RCW\,179,Gum\,48d & Centaurus\\
GMF 319.0-318.7 &	$55$			&$1.6$				&	2.9	     & ... & Centaurus\\
GMF 324.5-321.4	&	$>14$		&$0.21$				&   $<1.5$ & 		IRDC 321.71+0.07 & ...\\
GMF 324.5-321.4b&  $4.2$			&$0.21$				&   5.0	& 	IRDC 321.71+0.07 & ...\\
GMF 335.6-333.6	&	$>84$		&$31$				&	$<36.9$ & RCW106, G333 & Centaurus\\
GMF 335.6-333.6b&	$16$			&$2.2$				&	15.0	& ... & Centaurus\\
GMF 341.9-337.1	&	$>94$		&$5.7$				&	$<6.1$	& ... & ...\\
GMF 343.2-341.7	&	$>20$		&$0.85$				&	$<4.3$	& ... & ...\\
GMF 358.9-357.4	&	$>28$		&$2.6$  				&	$<9.3$	& IRDC G357 & Centaurus\\
\label{tab:filCandVelo2}
\end{tabular}
\tablefoot{Column 1: GMF identification; Col. 2: total gas mass of the GMF; 
Col. 3: Dense gas mass of the GMF; Col. 4: Dense gas mass fraction, estimated 
as the ratio between Col.2 and Col.3 ; Col. 5: known star-forming regions or 
molecular clouds along the GMF.; Col. 6: Spiral arm association, if any.}
\end{table*}

\subsection{Giant molecular filaments in the Galaxy}

\begin{figure}[h]
\centering
\resizebox{\hsize}{!}{\includegraphics{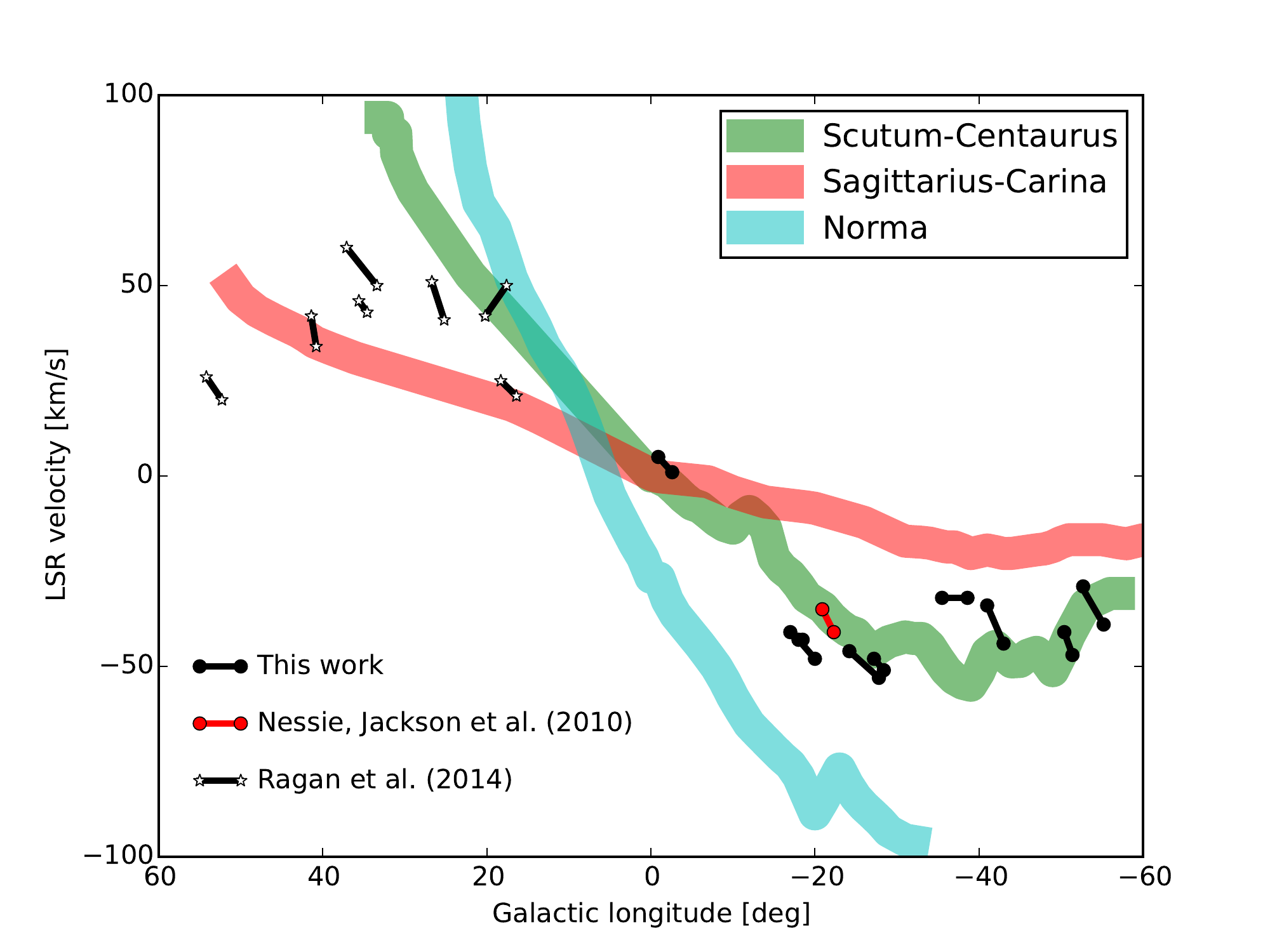}}
% distanceHisto.eps: 576x432 pixel, 72dpi, 20.32x15.24 cm, bb=0 0 576 432
\caption{LSR velocities of the Norma (cyan),
Scutum-Centaurus (green) and Sagittarius-Carina (red) spiral arms
as function of galactic longitude, as estimated by~\citet{reid14}.
The width of the lines, 8\,km/s, is equivalent to the spatial width
of the filaments from~\citep{reid14}. For simplicity, we only show the 
near kinetic distances of the spiral arms.
Each line segment represents a GMF, taking the
$v_{\mathrm{LSR}}$ values from the ends of the filaments. 
The line segments ended with black circles show GMFs
of our sample while those with white stars belong to R14.
We also show Nessie, with a red line ended in red circles.}
\label{fig:filVelo}
\end{figure}

What is the role of the GMFs in the Galactic spiral 
structure? Do they belong to the spiral arms or to inter-arm regions? 
We explore the answer to these questions in Fig.~\ref{fig:filVelo}, 
in which we show the $v_{\mathrm{LSR}}$
of the three inner spiral arms in the fourth quadrant 
as function of their galactic longitude following~\citet{reid14}. 
For simplicity, since our
filament finding method is more likely to reveal
filaments in the near side of the Galaxy (Sect.~\ref{sec:dist}),
we only show the near distance solutions of the spiral arms.
We measure the velocities at the ends 
of the filaments and plot them in Fig.~\ref{fig:filVelo}.
We find that six out of nine of our GMFs 
are connected to the Centaurus spiral arm, while 
three GMFs lie in inter-arm regions. 

We also re-examine the locations of the GMFs 
identified by R14 using the~\citet{reid14} model. 
We find that three out of their seven GMFs are connected to spiral 
arms (see Fig.~\ref{fig:filVelo}), two of them
to the Sagittarius arm and a third one to the Scutum arm. 
If we put together the GMF samples of R14 and this work
we find that nine out of 16 GMFs are related to spiral arms.
The percentage of spiral arm filaments higher
in the fourth quadrant (67\%) than in the first (<50\%).
In the first quadrant the GMFs are associated to the 
Sagittarius and Scutum arms, while in the fourth quadrant they are 
preferentially connected to the Centaurus arm.

The higher fraction of spiral arm filaments found by us compared
to R14 is affected by our use of the~\citet{reid14}
model instead of the~\citet{vallee08} model.
One of the main differences between~\citet{vallee08} and~\citep{reid14}
is that the latter compare their model
with typical spiral arm tracers such as \HII ~regions or massers
while this step is not done in the former work.
The velocities of the Sagittarius arm, the closest spiral arm in the first
quadrant, are significantly different 
in~\citet{vallee08} and~\citep{reid14} models.

\begin{figure}[h]
\centering
\resizebox{\hsize}{!}{\includegraphics{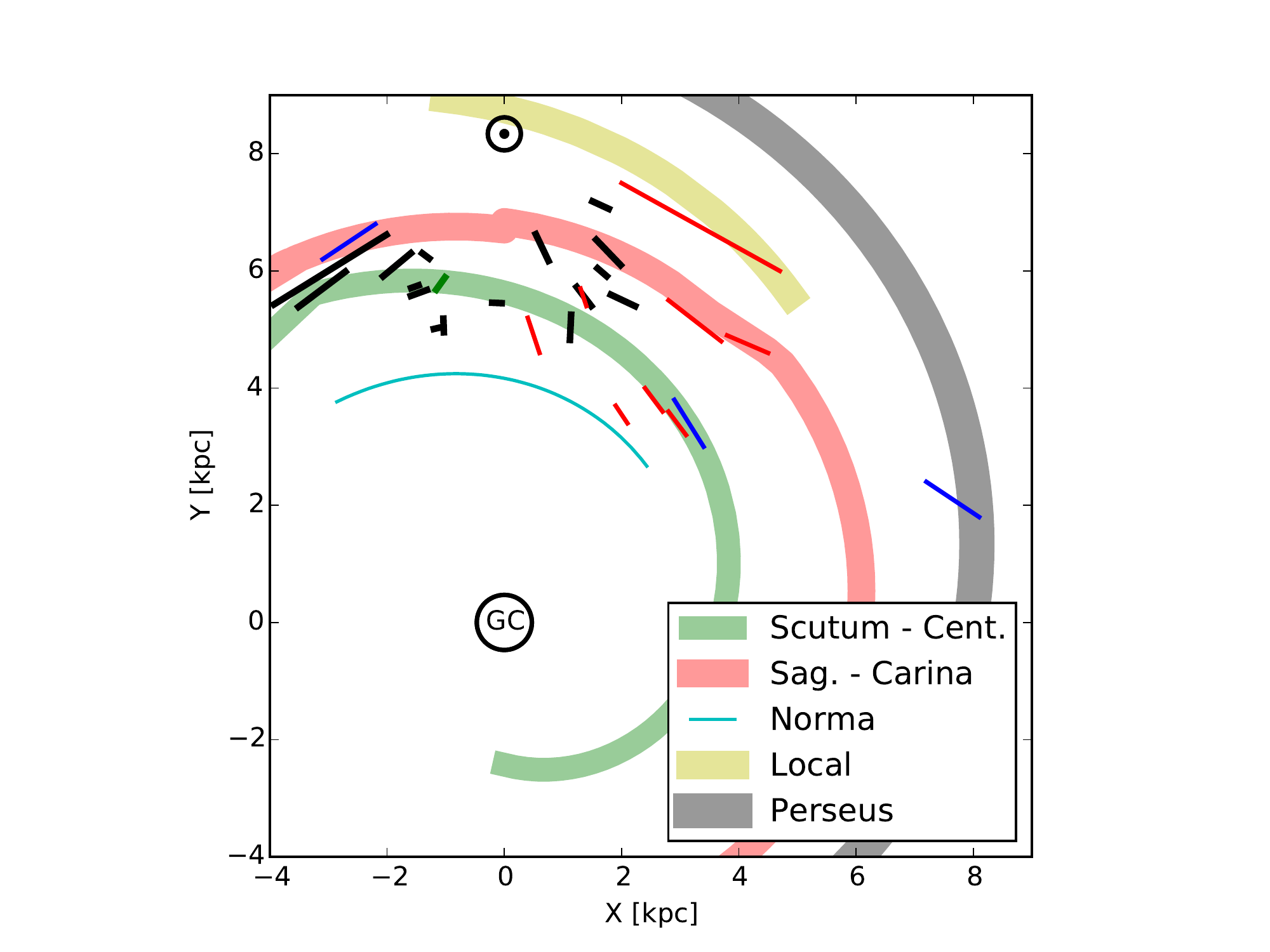}}
% distanceHisto.eps: 576x432 pixel, 72dpi, 20.32x15.24 cm, bb=0 0 576 432
\caption{Face-on view of the Milky Way spiral arm structure
	following the~\citet{reid14} model. We show the Norma (cyan), Scutum-Centaurus (green),
	Sagittarius-Carina (red), Local (yellow), and Perseus (Gray) spiral arms. The
	width of the spiral arms shows the width estimated by~\citet{reid14}.
	The circled GC represents the Galactic Center and $\sun$ represents the 
	Sun. The black lines represent the GMF samples of this work (at
	negative X values) and R14 (at positive X values). The red lines
	show the filament sample from~\citet{wang15}, based on \textit{Herchel}
	emission data. The green line shows Nessie~\citep{jackson10,goodman14}.
	The blue lines represent other previously known
	filaments~\citep{beuther11,battersby12,li13}.}
\label{fig:filFaceOn}
\end{figure}

Now we study the relative orientations of the GMFs 
compared to the spiral arms.   
Here we analyze together the GMF samples of this work and R14.
Figure~\ref{fig:filFaceOn} shows that there are seven 
GMFs connected to the Scutum-Centaurus arm, two to the 
Sagittarius arm and there are seven GMFs located in inter-arm regions.
Four of the GMFs 
connected to spiral arms in Fig.~\ref{fig:filFaceOn} are connected only by
one of their ends. We now address whether the uncertainties are large 
enough to place part of a filament
in an inter-arm region even if it lies within a spiral arm.

There are three main sources of error that play a role in 
the location of the filaments: the spiral arm width, the velocity resolution
of the \CO ~spectra, and the uncertainty in the kinematic distance. 
The latter is by far the most important. 
The distance uncertainties of our GMFs, obtained from the~\citet{reid14}
model, range between [0.7--0.3]\,kpc 
(see Table.~\ref{tab:filCandVelo}). These uncertainties are larger than 
the widths of the Scutum-Centaurus and Sagittarius spiral arms,
0.17\,kpc and 0.26\,kpc respectively, in the~\citet{reid14} model. 
The estimates of the exact location of the filaments, 
based on these kinetic distances alone, are not enough to claim
are not enough to claim that the GMFs are located 
in spiral arms, inter-arm regions or connect both. 

We now complement this study with 
PV diagrams of the \tCO ~emission, integrated between
$\lvert b \rvert \leq 1 \degr$.
If a GMF lies inside a spiral arm, its velocity should be 
consistent with that of the spiral arm. 
A spiral arm appears in a PV 
diagram as a strong single component. We overlay
a line showing the velocity of the GMF as function of its 
Galactic longitude on the PV diagram. If this line falls 
completely inside \tCO ~emission of any spiral arm, we could say
that the GMF is completely within it. If it falls outside, we
can confirm that the GMF is located in an inter-arm region.
We show the results of this experiment in the bottom
panels of Fig.~\ref{fig:apF3072_3054}--\ref{fig:apF3589_3574}.
These figures confirm that every GMF connected to a spiral arm in
Fig.~\ref{fig:filFaceOn} has velocities consistent with the arm over its
whole extent. 
These data offer no support for the hypothesis that some of the GMFs 
could be a spur (i.e., a filament connecting
spiral-arms with inter-arm regions observed in external
galaxies.)
Similarly, the inter-arm GMFs show velocities not consistent with
the spiral arms. 
We note that the spiral arm positions of some of our GMFs are 
independently confirmed by previous works, focused on particular \HII 
~regions or IRDCs within them (see App.~\ref{sec:describe}). 

\section{Discussion}

\subsection{Comparing large-scale filament finding methods}\label{sec:comp}

Three methods have been used so far to systematically
search for tens-of-parsec scale filaments. Two of them,
based on identification of the filaments as extinction features 
against the MIR/NIR background of the Galactic plane,
have been used by R14,~\citet{zucker15}, and this work.
In this work and in R14 we look for the largest 
filamentary structures in the Milky Way, irrespective of
their relative orientation with respect to the galactic midplane 
or spiral arms. The filaments revealed by both works are
known as GMFs. Instead,~\citet{zucker15} search explicitly
for Nessie analogues 
(i.e., filaments within spiral arms and parallel to the
Galactic midplane). They refer to these filaments bones.
The third method identifies the filaments as extended
emission features at far-infrared (FIR) wavelengths using \textit{Herschel} 
data~\citep{wang15}. We will refer to these as emission-identified filaments.
In principle, this naming scheme does not imply that the physical
properties of the objects are different, nor that they
should be called differently.

We compare first the GMFs and the bones. 
In the galactic plane areas common to~\citet{zucker15}, R14,
and this work there are nine bones, including Nessie, and 
16 GMFs, seven from R14 and nine from this paper.
At first look, the bones should be a sub-group of the 
GMFs in which only GMFs (or sub regions within them)
parallel to the Galactic mid-plane and inside spiral arms would be
identified as bones. However, only three bones correspond to
our GMFs. The filaments 10, 8 and 5 in~\citet{zucker15} are 
sub-regions of the GMF 335.6-333.6, GMF 358.9-357.4, and
GMF 20-17.9\footnote{As it has been noted by~\citet{zucker15} that
this GMF may not satisfy all the requirements to be considered as 
such. Concretely, it shows a steep jump in velocity in the PV
diagram.}, respectively. The main reasons why only three 
out of nine bones of~\citet{zucker15} overlap with the GMFs 
is because these bones have angular lengths clearly below $1\degr$, 
which is one of our GMF requirements. This property makes the bones 
and GMFs not directly comparable to each other. 

We now compare the extinction and emission-identified filaments.
Only three out of nine emission filaments in~\citet{wang15} 
have also been identified in extinction. These filaments,
labeled in~\citet{wang15} as G339, G11, and G26,
correspond respectively to Nessie,
the filament 6 in~\citet{zucker15}, and the GMF 26.7-25.4 in R14.
The emission-identified filaments can be missed as extinction filaments
because of lack of contrast between  the filament and the MIR/NIR 
background. On the other hand, extinction filaments may
not be identified in emission due to background and foreground 
confusion along the line-of-sight. Despite these differences both 
methods are likely to reveal quiescent structures in the 
early stages of star-formation. 
We conclude that the extinction and emission filament finding 
methods compliment each other well, finding filaments that 
can only be identified using one of both methods.

Do the physical properties of the filaments identified with the three 
different techniques agree?
The bones have lengths between 13\,pc and 52\,pc and are
the smallest of the three samples. The lengths of the emission-identified 
filaments (37--99\,pc) are comparable to those of the GMFs 
(see Table~\ref{tab:filCandVelo}). 
The bones and emission-identified filaments have
masses on the order $M=\sim10^{3}-10^{4}\,\mathrm{M_{\sun}}$,
and with the GMFs being the most massive large-scale filaments
(see Table~\ref{tab:filCandVelo2}). The masses of the emission-identified filaments
and the bones are comparable to the dense gas masses of the GMFs.
This is a selection effect.  
The masses of the bones are obtained over an area equivalent to the extinction
feature and not over the full \CO ~emission as it is the case for our 
GMFs\footnote{Not to mention that they are generally smaller
than the GMFs and the emission-identified filaments}.
The similarity between the masses of the emission-identified filaments and
the dense gas mass of the GMFs is a consequence of 
the dense gas material traced by the FIR continuum at 
350\,$\mu\mathrm{m}$ and 500\,$\mu\mathrm{m}$.
The area covered by this emission is limited to dense regions
that are surrounded by more diffuse \CO ~emission. 
The masses of the emission filaments and bones are therefore
comparable only to the dense gas mass of the GMFs rather
than to the total mass.

The direct comparison of the physical properties of the
different kind of large filaments found so far is not straightforward.
Each of the techniques used so far measures the filament properties on 
its on way. We encourage the use of the technique used in this paper to 
obtain the bulk properties of long filaments in the future,
so that they can be compared to the existing sample.

Do these filament techniques preferentially find spiral- 
or inter-arm filaments? In this discussion we do not
include the bones since they lie within spiral arms 
by definition.  Seven out of nine (78\%) emission-identified filaments lie in 
spiral arms~\citep{wang15}, as shown with the red filaments
of Fig.~\ref{fig:filFaceOn}. This percentage is
lower for the extinction filaments (11 out of 18, 61\%).
Although the percentage of spiral arm emission-identified 
filaments is larger than that of extinction filaments,
we acknowledge that with the small number of statistics 
we have this different may not be significant.

\subsection{Dense gas mass fraction and its variation with the filament location}

R14 found a tentative anti-correlation between the DGMF of the filaments and
their distance to the Galactic midplane. 
They also found that the filament with highest
DGMF in their sample was located in a spiral arm. This is also the case
of the large-scale filament with the highest DGMF known to date,
Nessie, 
with a DGMF$\sim50\%$~\citep{goodman14}.
However, we note that the dense gas mass estimates in~\citet{goodman14}
are made using HCN and not using sub-mm dust
emission as in R14 or this work. 
The results of R14 have however poor statistics due to the
low number of GMFs in their sample.
We explore this relationship further with our extended GMF sample.

\begin{figure}[t]
\centering
\resizebox{\hsize}{!}{\includegraphics{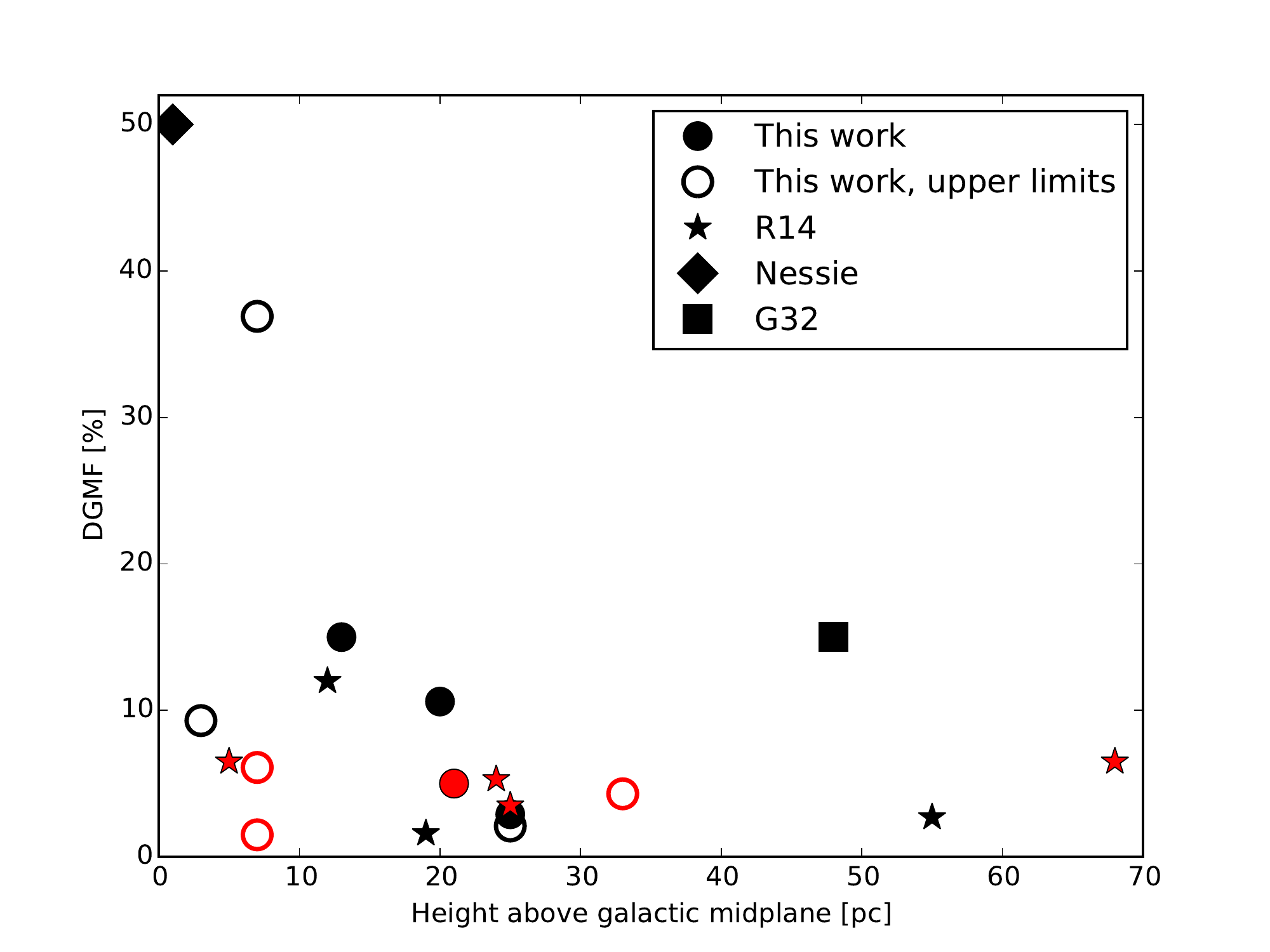}}
% distanceHisto.eps: 576x432 pixel, 72dpi, 20.32x15.24 cm, bb=0 0 576 432
\caption{Dense gas mass fraction of filaments as a function of the offset above the  
 	Galactic mid-plane. The full circles show our 
	GMF sample. The open circles show upper limits to 
	the DGMF in our GMF sample. 
	The black stars show the GMF sample of R14, the diamond
	indicates Nessie
	and the square the filament G32~\citep{battersby12}.
	The black symbols indicate GMFs within spiral arms while 
	the red ones show inter-arm GMFs.}
\label{fig:dgmf-z}
\end{figure}

The GMFs lie preferentially above the physical galactic midplane, which is
located at about $b=-0.35\degr$ in the fourth quadrant. 
In Fig.~\ref{fig:dgmf-z} we show the relationship between the height above
the galactic plane, $z$, and the DGMFs of the GMF samples of this work,
R14,~\citet{battersby12} and Nessie. It shows no correlation 
at all between $z$ and the DGMFs. The two filaments 
with the largest DGMFs lie close to the galactic midplane.
However, the scatter in the DGMFs at any $z$,
and particularly at $z<10$\,pc, is also very high. 
We therefore conclude that there is no evidence for correlation 
between the height above the galactic plane and the DGMF 
of the filaments. However, in this work we only cover $z$ scales
of a few tens of parsecs, while we are able to study 
kilo-parsec scales along the line of sight. We cannot 
rule out the possibility of a DGMF-$z$ relation at larger scales.

Another interesting points to look at are the difference in the DGMFs of
spiral arm filaments and inter-arm filaments and the
difference in DGMFs within different spiral arms. 
We found that the DGMF of the GMFs that belong to
the Scuttum-Crux arm is $13.0\pm10.1\,\%$, while that of the 
GMFs in the Sagittarius arm is $2.1\pm0.5\,\%$, where the uncertainties
are the standard deviation of the mean DGMFs. However,
only two GMFs lie in the Sagittarius arm, generating a very
poor statistical sample. 
Now we compare the DGMFs of arm and inter-arm GMFs.
We estimate the mean DGMFs for every giant filament (GMF or emission) with DGMF estimates.
We find that the mean DGMF in spiral arm filaments
is $14.3\pm15.5\,\%$ while this value is lower in the inter-arm
filaments, $4.8\pm1.7\,\%$. We note the big scatter in the DGMF of the
spiral arm filaments, due mainly to Nessie and GMF 335.6-333.6, 
with DGMFs of 50\% and 37\% respectively. 
We performed an independent-samples t-test 
to compare the DGMFs of spiral- and inter-arm GMFs.
We used the task \texttt{ttest-ind} from the package \texttt{scipy} in
\texttt{Python}. This task returns the $t$-value and also a $p$-value\footnote{
If the $p$-value returned by \texttt{ttest-ind} is lower than 0.10,
then both means are significantly different.} 
that is an indication on the significance of the means. 
We found a significant difference in the DGMFs 
of spiral-arm ($14.3\pm15.5\,\%$) and inter-arm GMFS
($4.8\pm1.7\,\%$), with $t$=2.02 and $p$=0.07.
We can therefore assume that the mean DGMF of the spiral 
arm GMFs is higher than that of the inter-arm filaments. 
This is in agreement with observations in external galaxies, where
the amount of dense gas is larger in the spiral arms than in 
inter-arm regions~\citep{hughes13,schinnerer13}.
The connection between the DGMF and the star-forming activity of 
molecular clouds~\citep{kainul09,lada10,lada12,abreu15} and this result suggest
that the spiral arm filaments have larger star-forming potential than the 
inter-arm filaments. In other words, the star-forming activity of the
GMFs depend on its Galactic location with respect to the spiral arms. 

The masses of our GMFs, and also those of R14, are consistent
with the definition of giant molecular clouds (GMCs).~\citet{stark06}
found, using a sample of 56 GMCs (defined by them as molecular 
clouds with $M>10^{5}$\,M$_{\sun}$), that all GMCs were related 
to spiral arms. Only a 10\% of less massive clouds were found
to be unrelated to spiral arms. Following the GMC definition of~\citet{stark06}, 
we found that five out of 14 GMFs consistent with GMC masses are 
in inter-arm regions. We therefore find that 
most of the GMCs in the Galaxy are related to spiral arms, 
as it was found by~\citep{dame86}. 
Although the GMC population is enhanced in the spiral arms,
the star-forming activity is not significantly enhanced on them~\citep{eden12,eden13,moore12}.
Our results agree with a picture on which a non-negligible amount GMCs 
can be found outside spiral arms, as also seen in external 
galaxies~\citep{schinnerer13}. Also in external galaxies,~\citet{foyle10}
have reported significant star-forming activity in inter-arm regions.

\section{Conclusions}

We have used the 2MASS, GLIMPSE, and ThruMMS surveys to
extend the GMF catalog initiated in R14 to the fourth Galactic quadrant. 
We inspected visually the NIR/MIR images to look for 
filamentary extinction features of at least one degree
in angular length. We then used spectral 
\CO ~information from the ThruMMS survey 
to confirm that those features are continuous in velocity.

\begin{itemize}

\item We present a sample of nine newly identified GMFs. 
The projected lengths of the new GMFs range
from 38\,pc to 168\,pc. 
Total masses as traced by \CO ~are between $4\times10^{4}\,\mathrm{M_{\sun}}$
and $9.4\times10^{5}\,\mathrm{M_{\sun}}$. 
We also used the cold dust emission at 870\,$\mu\mathrm{m}$ to estimate the
dense gas mass of the GMFs and found that it ranges
from $2.1\times10^{3}\,\mathrm{M_{\sun}}$ to
$3.1\times10^{5}\,\mathrm{M_{\sun}}$. 

\item The ratio of the dense and total gas masses
is the DGMF, which ranges between 1.5\% and 37\%. The largest 
is related with the \HII ~complex G333 in the GMF 335.6-333.6. 
This value agrees with the DGMFs of massive 
(>$10^{5}\,\mathrm{M_{\sun}}$) molecular
clouds with \HII ~regions found in~\citet{abreu15}. The other values, between
1.5\% and 15\%, are consistent with typical DGMFs 
found in molecular clouds~\citep{battisti14,abreu15}.

\item We explored the role of the GMFs identified
by us and R14 in the Galactic context. 
Adopting the~\citet{reid14} Galactic model, 
we find that nine out of 16 GMFs are connected 
to spiral arms. Seven out of these nine filaments are connected to the
Scutum-Centaurus arm and two to the Sagittarius arm.
Three GMFs of R14 are related to 
spiral arms when the~\citet{reid14} model is used, 
while only one is if~\citet{vallee08} model is used.

\item We find no correlation between the DGMFs of
GMFs and the distance from the Galactic midplane.
This result disagrees with the tentative correlation
found by R14. However, we note that we only 
observed GMFs within a few tens of parsecs of the Galactic midplane.
We found that the DGMFs of the spiral
arm GMFs are larger than those of the inter-arm 
GMFs. This result agrees with observations of external 
galaxies showing that the DGMFs of molecular clouds 
within spiral arms have larger DGMFs than inter-arm clouds
~\citep{hughes13,schinnerer13}. 
The DGMF has a direct relationship
with the star-forming activity~\citep{kainul09,lada12}. 
This result therefore suggests
that the star-forming potential of the GMFs is tightly 
connected to their relative position to the Galactic
spiral arms. 

\item We compared the different methods used to date to 
identify large filaments: filaments identified as
extinction features (GMFs and bones)
and emission-identified filaments. 
The GMFs and the emission-identified filaments have comparable sizes 
and are generally larger than the bones. 
The total masses of the bones
and the emission filaments are comparable to the 
dense gas masses of our GMF sample. This result is an observational
effect since both, bones and emission filaments, search 
preferentially dense gas. Emission filaments are more 
preferentially connected to the spiral arms than 
our GMFs. Due to the different biases of the extinction
and emission filament finding methods, each method can
identify filaments that are missed by the other. 

\end{itemize}
\begin{acknowledgements}
The work of J.A. is supported by the Sonderforschungsbereich (SFB) 881 
\textquotedblleft The Milky Way System \textquotedblright and
the International Max-Planck Research School (IMPRS) at Heidelberg University.
The work of J.K. was supported by the
Deutsche Forschungsgemeinschaft priority
program 1573 (\textquotedblleft Physics of the Interstellar
Medium\textquotedblright). 
S.E.R. acknowledges support from VIALACTEA, 
a Collaborative Project under Framework Programme 7 
of the European Union, funded under Contract \# 607380.
We thank to M. Reid and T. Dame for providing us with
an updated version of their Galactic model and also
for very useful discussions that helped to improve 
the manuscript.
This paper has used information
from the RMS survey database at www.ast.leeds.ac.uk/RMS 
constructed with support from the Science and Technology 
Facilities Council of the UK. This research has made use of the SIMBAD 
database, operated at the CDS, France.
\end{acknowledgements}

% for the bibliography, at the end
\bibliographystyle{aa} % style aa.bst
\bibliography{bibliography} % your references Yourfile.bib
\appendix

\section{Notes on individual GMFs}\label{sec:describe}

\subsection{GMF 307.2-305.4}\label{sec:F305}

\begin{figure*}
\centering
\includegraphics[width = 0.95\textwidth]{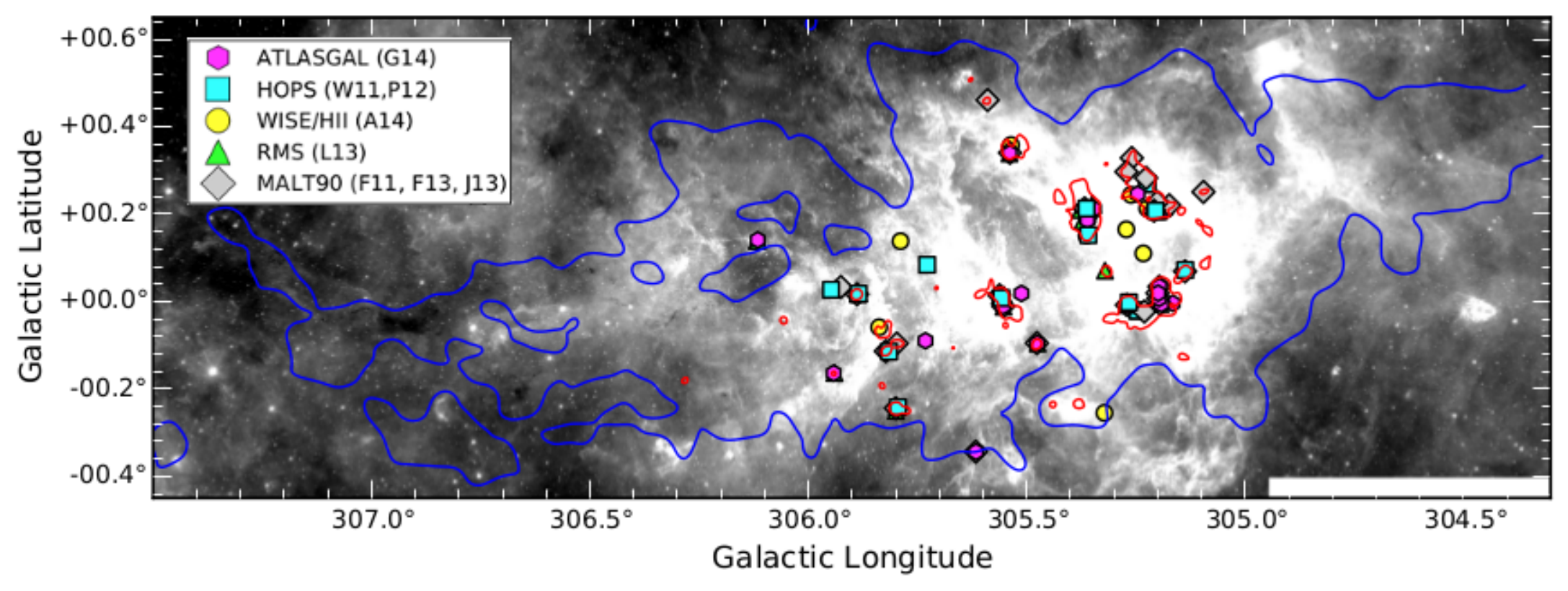}
\includegraphics[scale = 0.4]{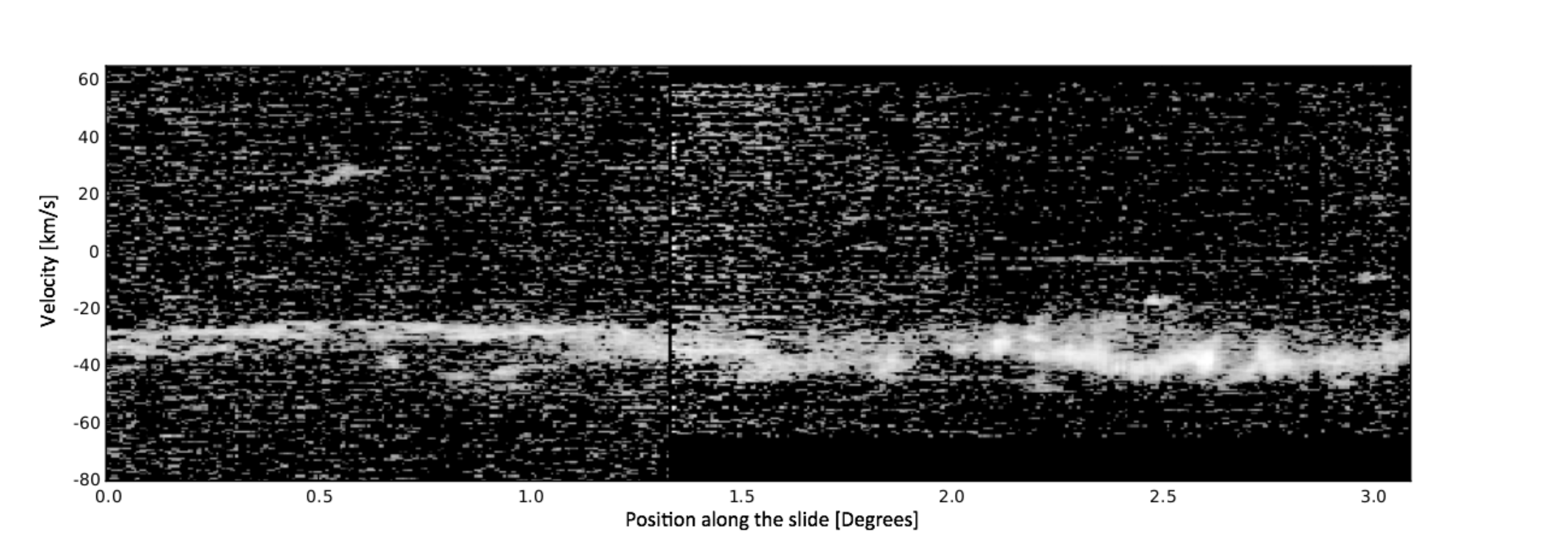}
\includegraphics[scale = 0.4]{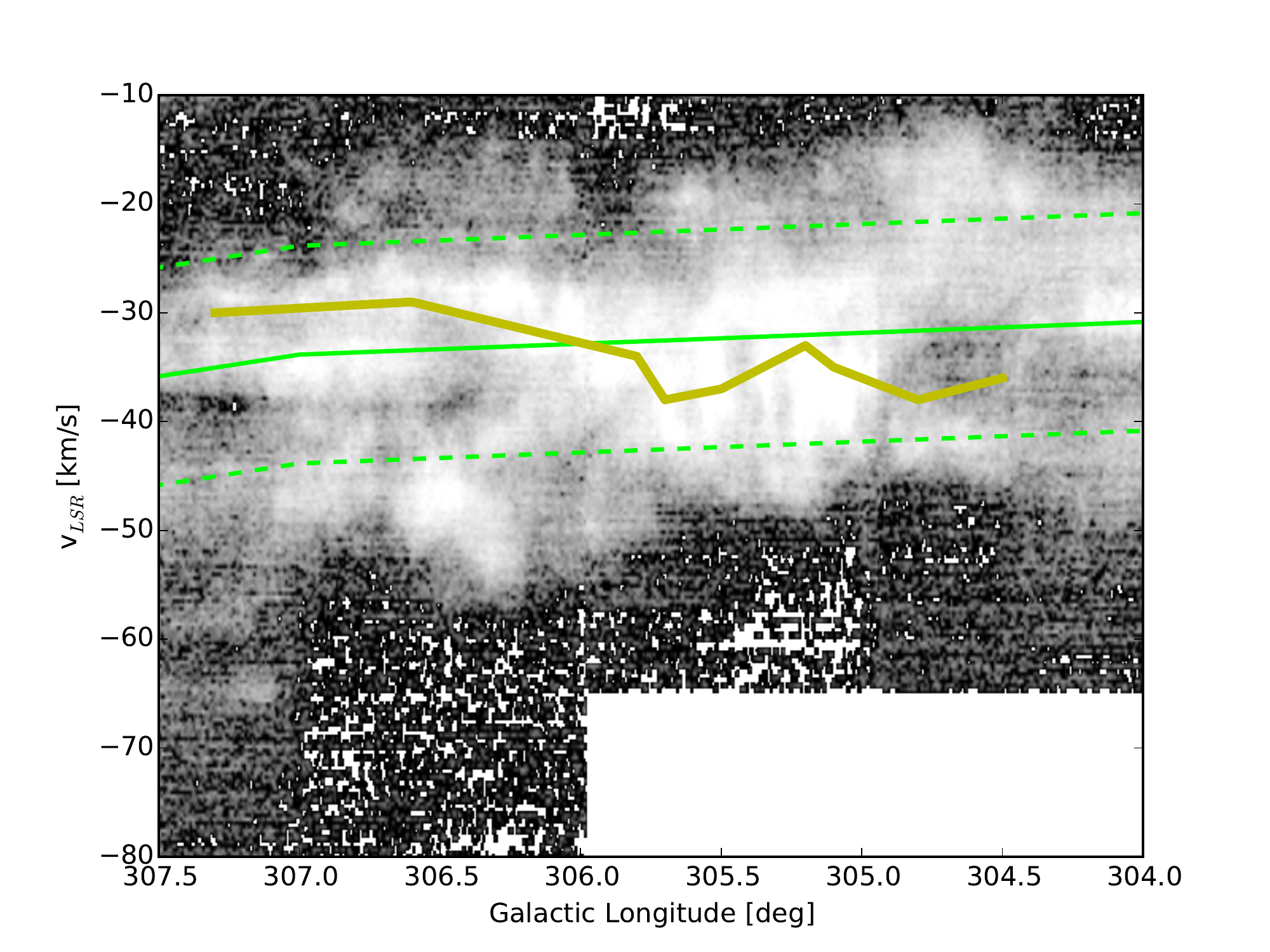}
\caption{\textit{Top:} Grayscale GLIMPSE $8\,\mu$m image of the GMF 307.2-305.4.
The blue contours show the \CO ~integrated intensity of 2\,K\,/km/s,
integrated over the velocity range [-45,-25]\,km/s.
The red contours show the ATLASGAL emission at a contour level of 
$F_{870\mathrm{\,\mu m}}=250$\,mJy/beam. 
The filled geometric objects show all the dense gas measurements from 
different surveys with $v_{\mathrm{LSR}}$ within the velocity range  
indicated in Table~\ref{tab:filCandVelo}. 
Gray diamonds show MALT90~\citep{foster11,foster13,jackson13}
survey measurements, yellow circles belong to the spectral catalog
of \HII ~regions in the WISE survey~\citep{anderson14}, 
The RMS survey~\citep{lumsden13} is represented by green triangles, the cyan
squares show NH$_{3}$ clumps from HOPS survey~\citep{purcell12}, and
the pink hexagons show spectral follow-ups of the ATLASGAL 
survey~\citep{giannetti14,urquhart13a,urquhart14b}.
\textit{Middle:} Position-velocity diagram of the \CO ~line of the 
GMF 307.2-305.4, obtained
 from a slice following the extinction feature (black line in top panel) used to identify 
 GMF 307.2-305.4. \textit{Bottom:}
PV diagram of the \tCO ~emission between $\lvert b \rvert \leq 1\degr$.
The yellow line shows GMF 307.2-305.4 in the PV space.
The green solid line shows the Scutum-Centaurus arm as predicted by~\citet{reid14}
and the dashed green lines show $\pm10$\,km/s of the velocity of the spiral arm.}
\label{fig:apF3072_3054}
\end{figure*}

The GMF 307.2-305.4 can be identified as a mixture of emission and 
extinction features (see Fig.~\ref{fig:apF3072_3054}).
This filament shows the widest velocity range of the sample 
[-29,-39]\,km/s. This wide velocity spread can be caused by 
the location of the filament, close to the tangent point 
(see also Fig.~\ref{fig:filFaceOn}), or by 
the expanding bubble generated by the massive G305 \HII 
~complex~\citep{hindson10,davies12}, that can be 
identified as emission in the 8\,$\mu$m image. 
This \HII region has a molecular gas mass of 
$\sim6\times10^{5}_{\sun}$~\citep{hindson10}. 
We find several dense gas clumps in this
complex that have velocities consistent with the GMF.
GMF 307.2-305.4 lies within the Scutum-Centaurus spiral arm.
This is consistent with previous works that relate
G\,305 to the Scutum-Centaurus arm~\citep{hindson10,davies12}. 

\begin{table*}
\caption{L--B--V tracks of the filaments} %
%\label{Table1} % is used to refer this Table in the text
\centering % used for centering Table
\begin{tabular}{cccccccc} % centered columns (4 columns)
\hline\hline % inserts double horizontal lines
&&&&GMF\,307.2-305.4&&&\\\hline
l $[\degr]$&307.3&306.6&305.8&305.7&305.5&305.2&305.1\\
b $[\degr]$&0.14&-0.12&-0.10&-0.03&-0.03&-0.03&-0.05\\
v [km/s]   &-30&-29&-34&-38&-37&-33&-35\\\hline
&&&&GMF\,309.5-308.7&&&\\\hline
l $[\degr]$&309.2&309.1&309.0&308.7&&&\\
b $[\degr]$&-0.48&-0.16&-0.13&0.63&&&\\
v [km/s]   &-45&-43&-44&-46&&&\\\hline
&&&&GMF\,319.0-318.7&&&\\\hline
l $[\degr]$&319.3&318.8&318.5&318.3&318.1&317.7&317.5\\
b $[\degr]$&-0.08&-0.17&-0.23&-0.38&-0.07&0.07\\
v [km/s]   &-37&-38&-40&-38&-43&-40&-44\\\hline
&&&&GMF\,324.5-321.4&&&\\\hline
l $[\degr]$&323.9&321.5&&&&&\\
b $[\degr]$&-0.46&0.10&&&&&\\
v [km/s]   &-32&-32&&&&&\\\hline
&&&&GMF\,335.6-333.6&&&\\\hline
l $[\degr]$&335.2&334.6&332.9&332.3&&&\\
b $[\degr]$&-0.26&-0.21&-0.50&-0.48&&&\\
v [km/s]   &-41&-47&-54&-55&&&\\\hline
&&&&GMF\,335.6-333.6b&&&\\\hline
l $[\degr]$&332.7&332.5&332.3&331.9&331.6&331.4&\\
b $[\degr]$&-0.23&-0.13&-0.12&-0.11&-0.24&-0.33&\\
v [km/s]   &-45&-47&-50&-50&-46&-48&\\\hline
&&&&GMF\,341.9-337.1&&&\\\hline
l $[\degr]$&342.2&341.5&340.8&340.3&&&\\
b $[\degr]$&-0.13&-0.29&-0.23&-0.14&&&\\
v [km/s]   &-41&-41&-46&-45&&&\\\hline
&&&&GMF\,343.2-341.7&&&\\\hline
l $[\degr]$&342.8&342.1&341.9&341.8&&&\\
b $[\degr]$&0.07&0.19&0.29&&&&\\
v [km/s]   &-42&-41&-43&-43&&&\\\hline
&&&&GMF\,358.9-357.4&&&\\\hline
l $[\degr]$&358.4&358.1&357.8&357.5&357.1&&\\
b $[\degr]$&-0.48&-0.44&-0.32&-0.32&-0.02&&\\
v [km/s]   &4&5&7&5&7&&\\\hline

\label{tab:lbvtracks}
\end{tabular}
\end{table*}

\subsection{GMF 309.5-308.7}\label{sec:F309}

\begin{figure*}
\centering
\includegraphics[scale = 0.6,angle=270]{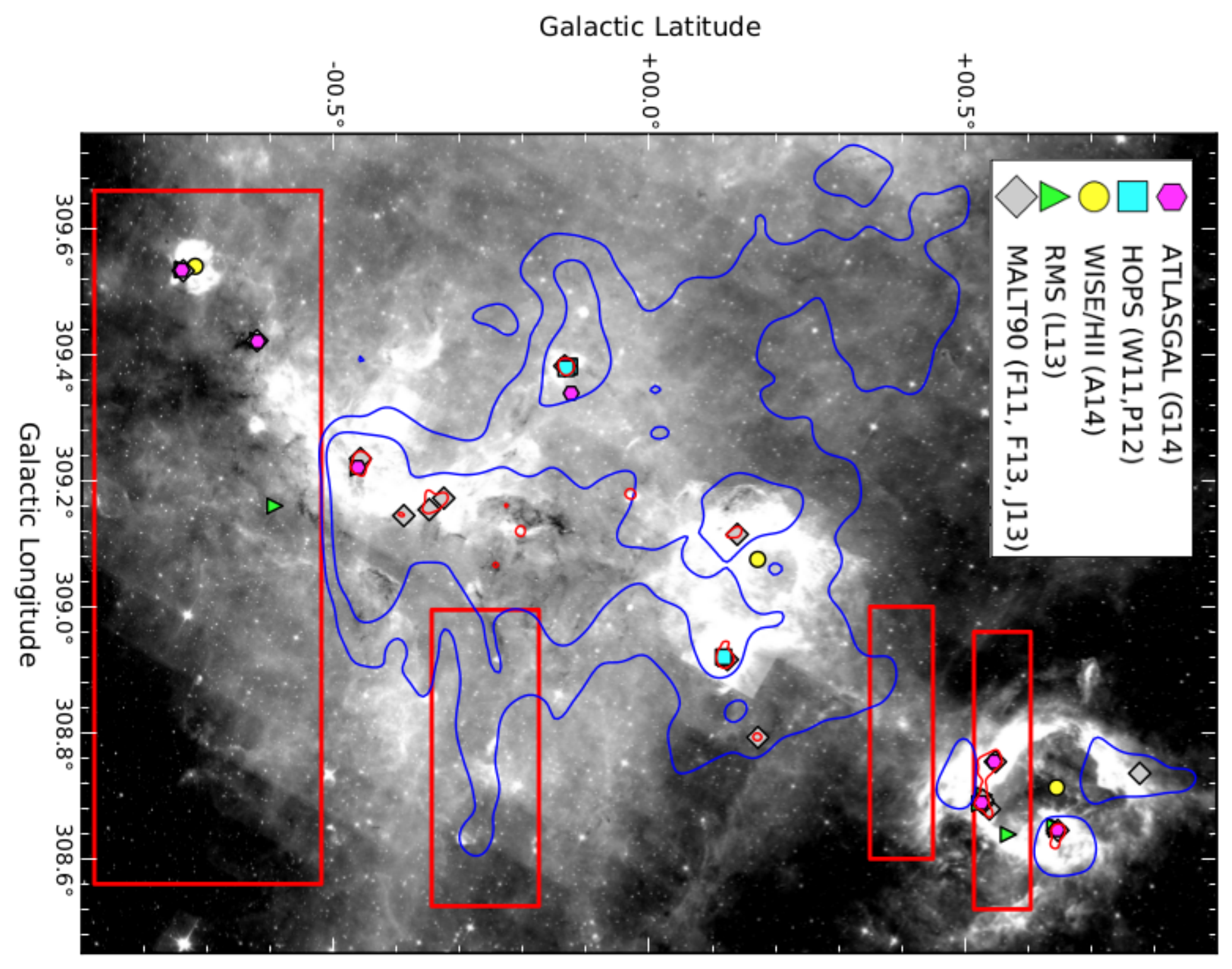}
\includegraphics[scale = 0.4]{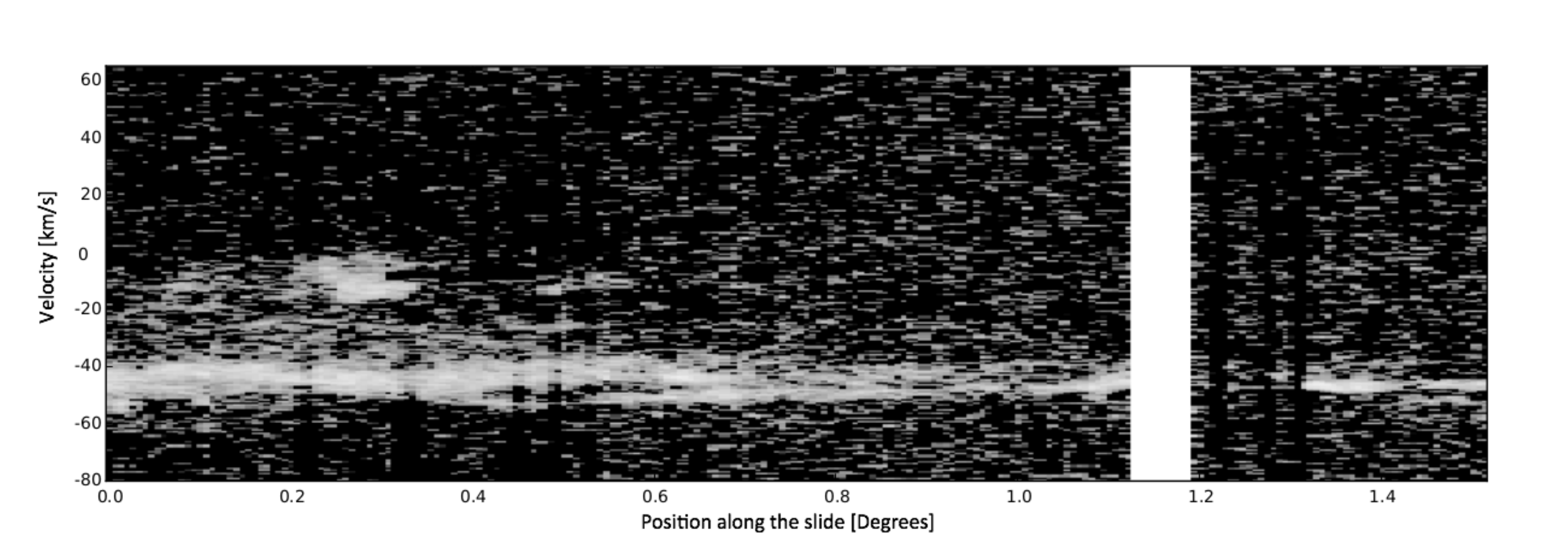}
\includegraphics[scale = 0.4]{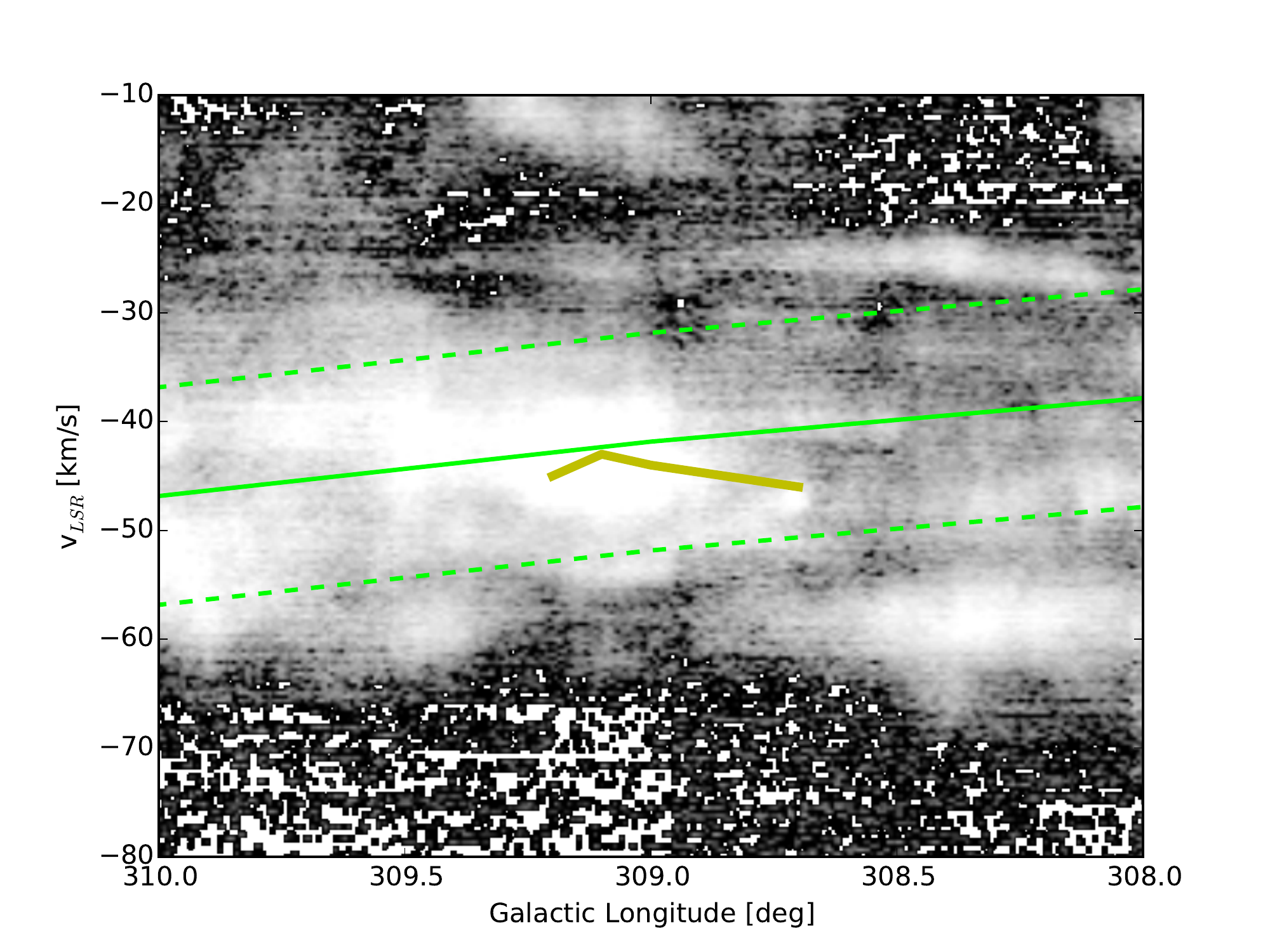}
\caption{\textit{Top:} Grayscale GLIMPSE $8\,\mu$m image of the GMF 309.5-308.7.
The blue contours show the \CO ~integrated intensity of 1.5\,K\,/km/s
and 3\,K\,/km/s,
integrated over the velocity range [-53,-35]\,km/s.
The red contours show the ATLASGAL emission at a contour level of 
$F_{870\mathrm{\,\mu m}}=250$\,mJy/beam. 
The red boxes show regions with poor ThruMMS data or absence of it.
The filled geometric objects show all the dense gas measurements from 
different surveys with $v_{\mathrm{LSR}}$ within the velocity range  
indicated in Table~\ref{tab:filCandVelo}. Symbols as in Fig.~\ref{fig:apF3072_3054}.
\textit{Middle:} Position-velocity diagram of the \CO ~line of the GMF 309.5--308.7, obtained
 from a slice following the extinction feature used to identify 
 GMF 309.5--308.7. \textit{Bottom:}
PV diagram of the \tCO ~emission between $\lvert b \rvert \leq 1\degr$.
The yellow line shows GMF 309.5--308.7 in the PV space.
The green solid line shows the Scutum-Centaurus arm as predicted by~\citet{reid14}
and the dashed green lines show $\pm10$\,km/s of the velocity of the spiral arm.}
\label{fig:apF3095_3087}
\end{figure*}

The only GMF aligned perpendicularly to the Galactic plane, 
it can be seen at 8\,$\mu$m as a vertical extinction feature connecting
a series of strong emitting regions, known to be
regions of massive star-forming activity:
RCW79~\citep{saito01,russeil03,zavagno06} in the north,  
and Gum 48d~\citep{karr09} in the south, the latter 
connected to the Scutum-Centaurus arm. 
Previous distance estimates of these \HII ~regions agree with those
found in this paper. Some supernova remnants
are found within GMF 309.5-308.7 and close to
it at lower longitudes, confirming the very active recent
massive star-forming activity in the region.  
Unfortunately, ThruMMS does not fully cover the filament.
The southern region has not been observed in either \tCO ~or \CO 
~(see red boxes in Fig.~\ref{fig:apF3095_3087}). The total molecular mass of the
GMF is therefore a lower limit.

\subsection{GMF 319.0-318.7}

\begin{figure*}
\centering
\includegraphics[width = 0.8\textwidth]{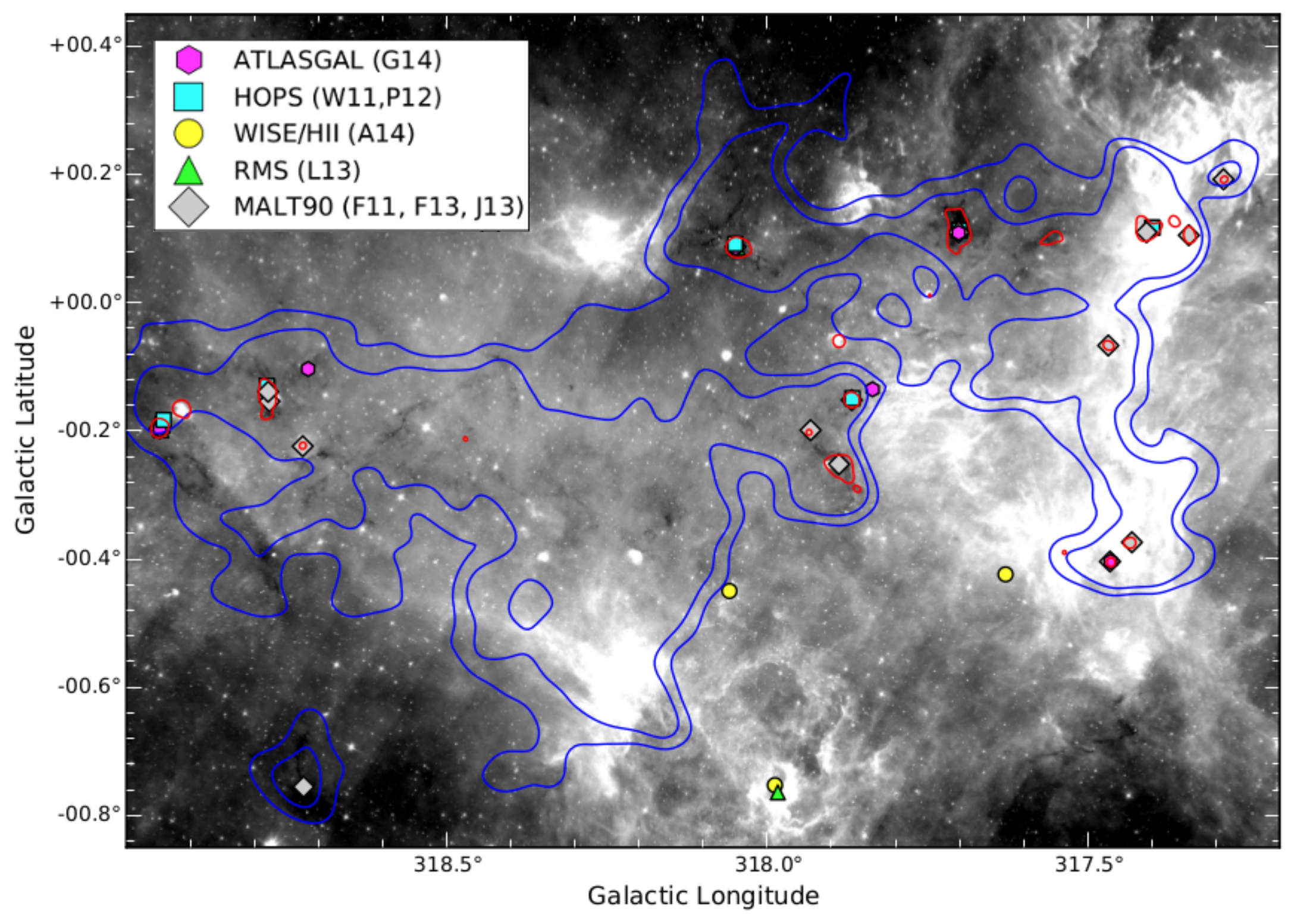}
\includegraphics[scale = 0.4]{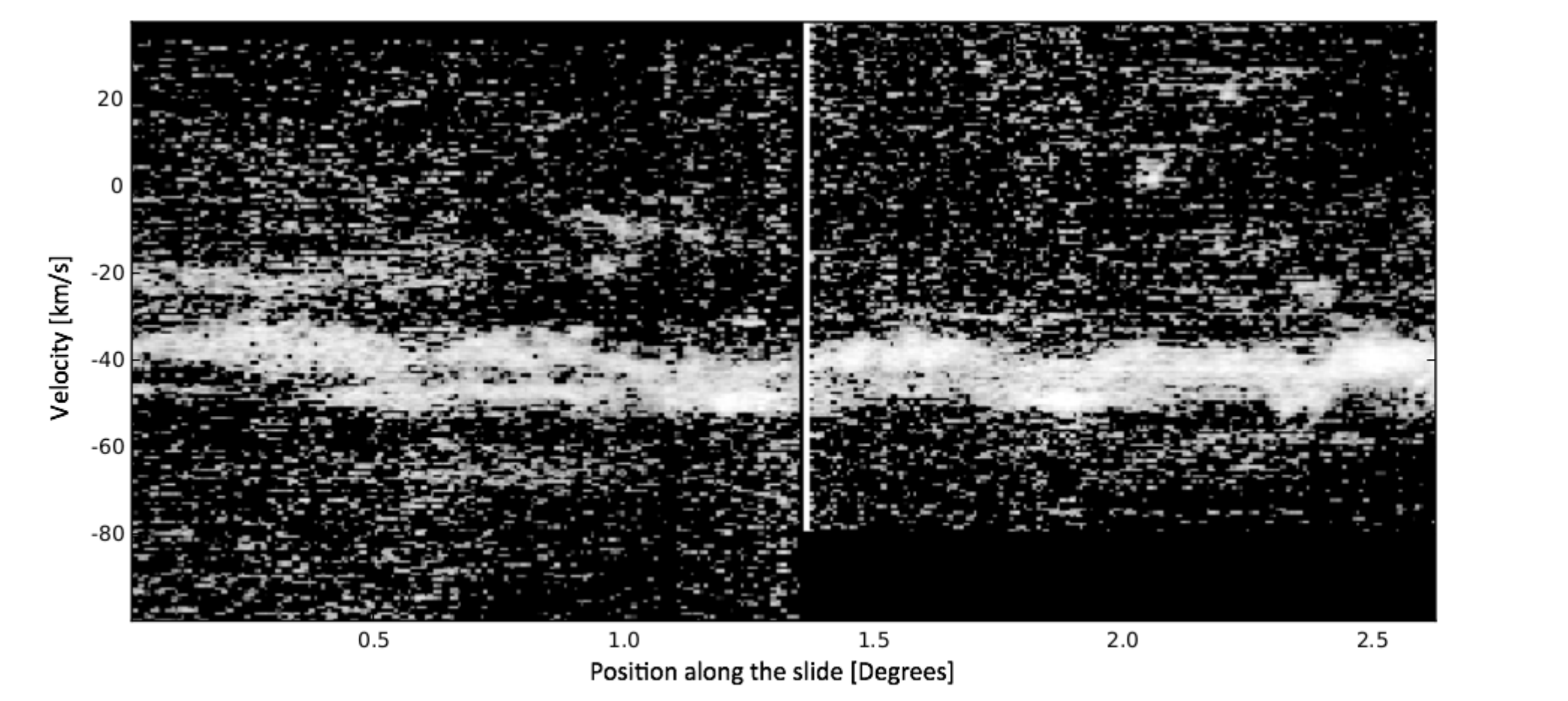}
\includegraphics[scale = 0.4]{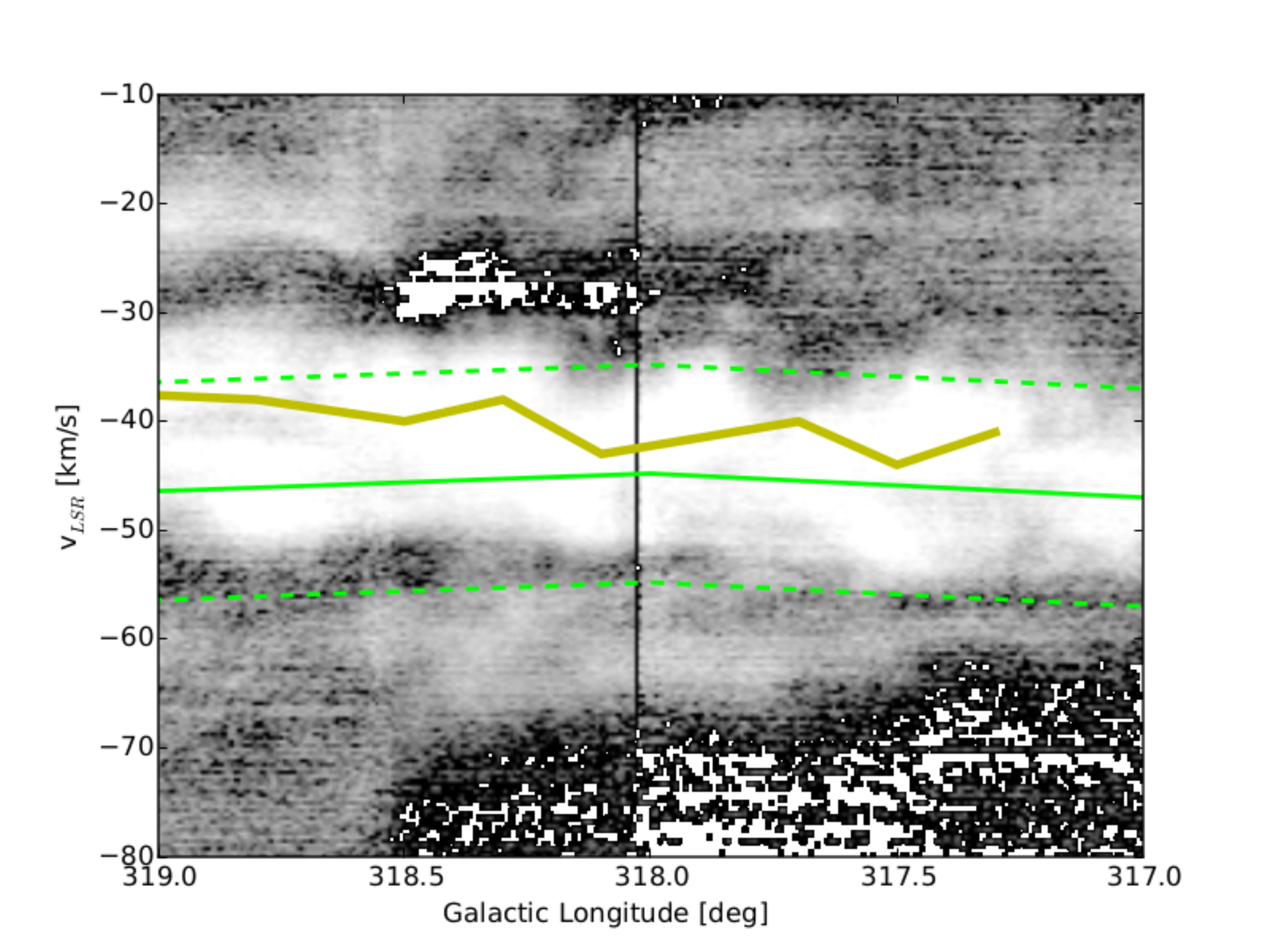}
\caption{\textit{Top:} Grayscale GLIMPSE $8\,\mu$m image of the GMF 319.0--318.7.
The blue contours show the \CO ~integrated intensity of 1.5\,K\,/km/s
and 2\,K\,/km/s,
integrated over the velocity range [-53,-34]\,km/s.
The red contours show the ATLASGAL emission at a contour level of 
$F_{870\mathrm{\,\mu m}}=250$\,mJy/beam. 
The filled geometric objects show all the dense gas measurements from 
different surveys with $v_{\mathrm{LSR}}$ within the velocity range  
indicated in Table~\ref{tab:filCandVelo}. Symbols as in Fig.~\ref{fig:apF3072_3054}.
\textit{Middle:} Position-velocity diagram of the \CO ~line of the GMF GMF 319.0--318.7, obtained
 from a slice following the extinction feature used to identify 
 GMF 319.0--318.7. \textit{Bottom:}
PV diagram of the \tCO ~emission between $\lvert b \rvert \leq 1\degr$.
The yellow line shows GMF 319.0--318.7 in the PV space.
The green solid line shows the Scutum-Centaurus arm as predicted by~\citet{reid14}
and the dashed green lines show $\pm10$\,km/s of the velocity of the spiral arm.}
\label{fig:apF3190_3187}
\end{figure*}

We identified the GMF as an extinction feature connecting two 
star-forming sites. However, a close look to the GMF presents it as two 
dense filaments, both following extinction features
connected by a diffuse envelope (see Fig.~\ref{fig:apF3190_3187}).
This GMF is located within the predicted Scutum-Centaurus arm.

\subsection{GMF 324.5-321.4}\label{sec:F323}

\begin{figure*}
\centering
\includegraphics[width = 0.9\textwidth]{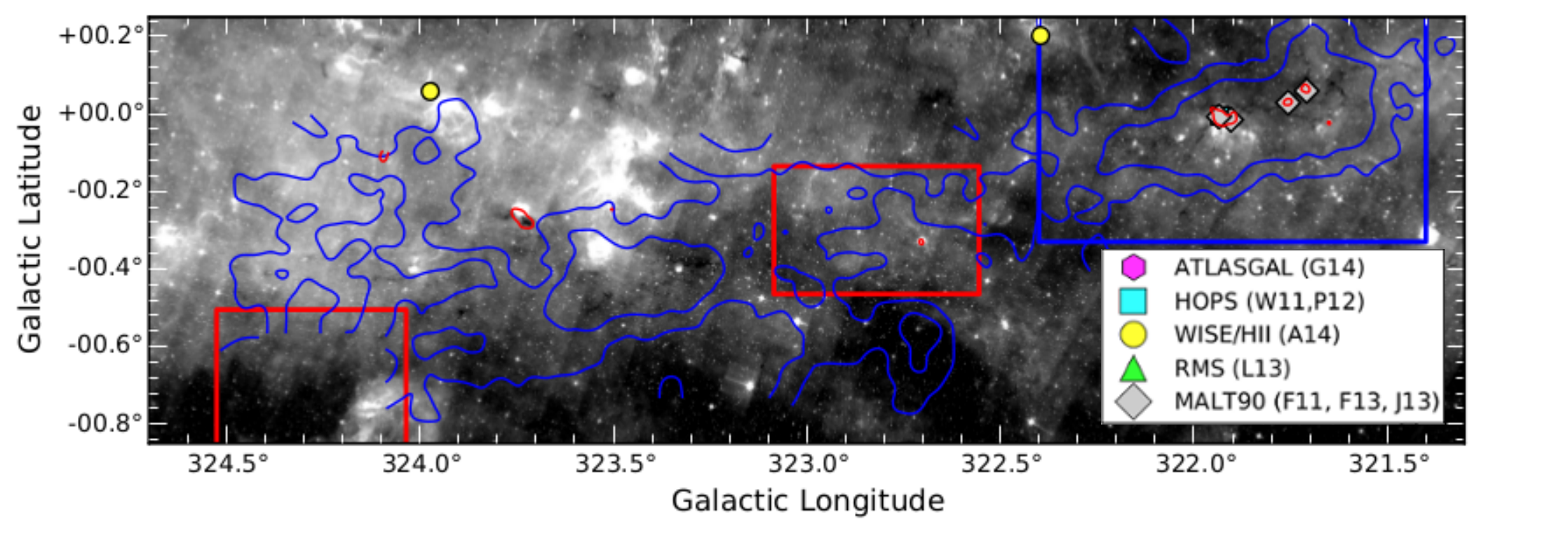}
\includegraphics[scale = 0.4]{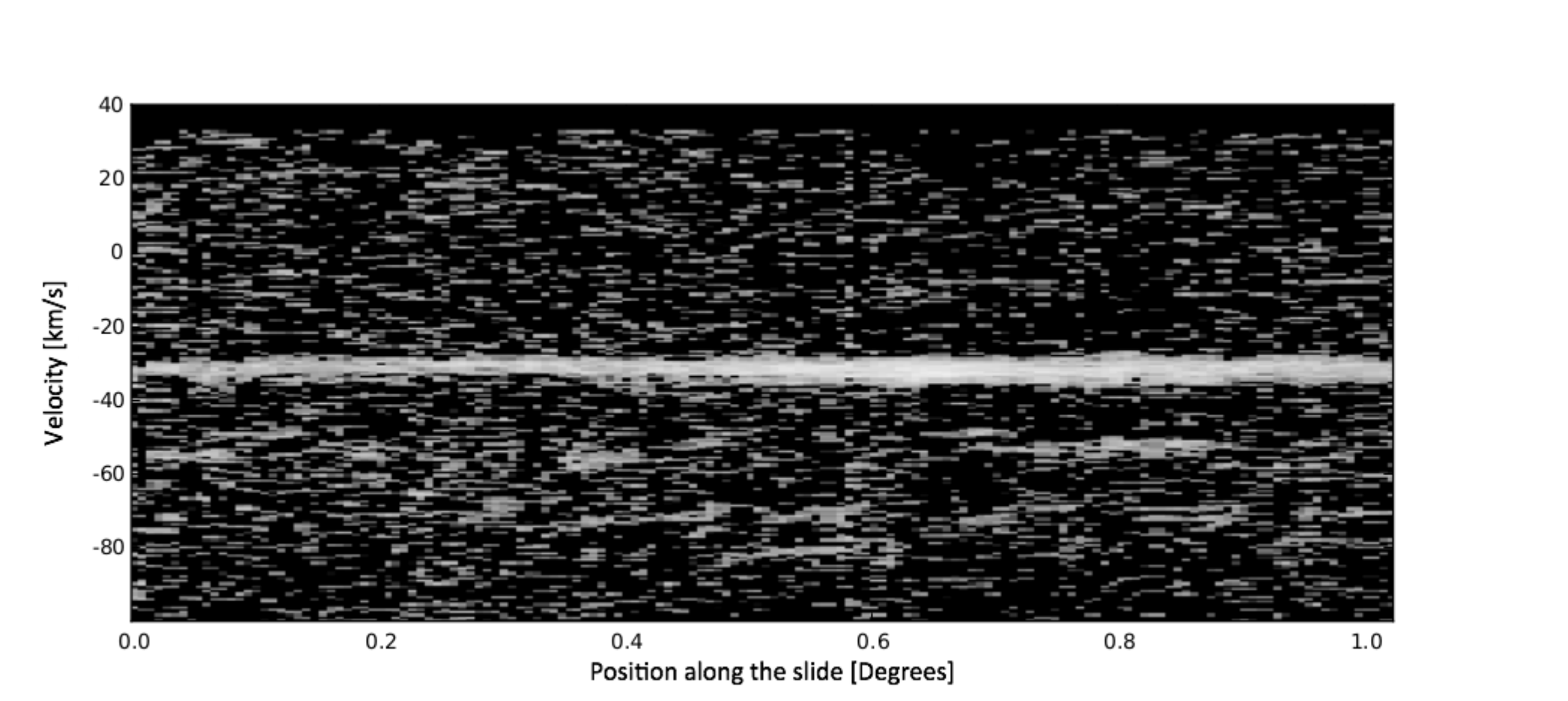}
\includegraphics[scale = 0.4]{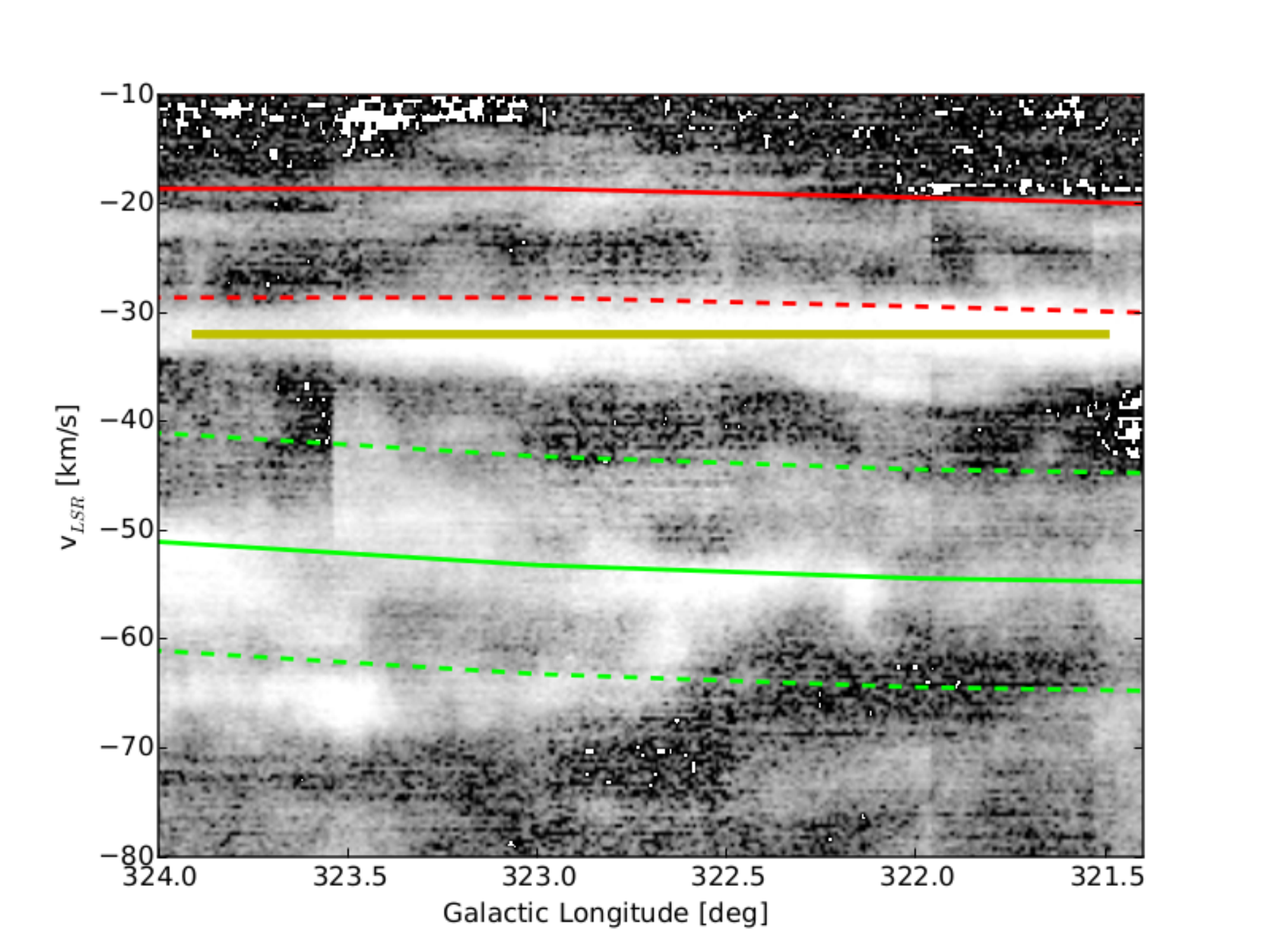}
\caption{\textit{Top:} Grayscale GLIMPSE $8\,\mu$m image of the GMF 324.5--321.4.
The blue contours show the \CO ~integrated intensity of 1.5\,K\,/km/s
and 2\,K\,/km/s,
integrated over the velocity range [-35,-28]\,km/s.
The red contours show the ATLASGAL emission at a contour level of 
$F_{870\mathrm{\,\mu m}}=250$\,mJy/beam. 
The red boxes show regions with poor ThruMMS data or absence of it
and the blue box shows GMF F3245--3214b. 
The filled geometric objects show all the dense gas measurements from 
different surveys with $v_{\mathrm{LSR}}$ within the velocity range  
indicated in Table~\ref{tab:filCandVelo}. Symbols as in Fig.~\ref{fig:apF3072_3054}.
\textit{Middle:} Position-velocity diagram of the \CO ~line of the GMF 324.5--321.4, obtained
 from a slice following the extinction feature used to identify 
 GMF 324.5--321.4. \textit{Bottom:}
PV diagram of the \tCO ~emission between $\lvert b \rvert \leq 1\degr$.
The yellow line shows GMF 324.5--321.4 in the PV space.
The green solid line shows the Scutum-Centaurus arm as predicted by~\citet{reid14}
and the dashed green lines show $\pm10$\,km/s of the velocity of the spiral arm.
The red solid line shows the predicted Sagittarius spiral arm and the red dashed 
line the --10\,km/s velocity of Sagittarius.}
\label{fig:apF3245_3214}
\end{figure*}

The most prominent extinction feature 
is the IRDC 321.71+0.07~\citep{vasyunina09}, located at
a distance of 2.14kpc.
The red box in the bottom-left corner of Fig.~\ref{fig:apF3245_3214}
shows that there is no ThruMMS coverage of that area. 
The red box in the middle of the filament shows a region with very high noise. 
In this region the east and west ends of GMF 324.5-321.4 are barely connected.
Although the \tCO ~map shows a clear connection between both parts
of the filament, we proceed with caution, dividing this GMF in two possible filaments:
the whole filament from $l=321.5\,\degr$ to $l=324.5\,\degr$, and a shorter version from
$l=321.5\,\degr$ to $l=322.5\,\degr$ called GMF 324.5-321.4b. 
All the dense gas mass of this filament is located in IRDC 321.71+0.07,
so that the DGMF of GMF 324.5-321.4b is considerably larger than that of the
longer GMF 324.5-321.4. This is the only filament showing no velocity
gradient along it. GMF 324.5-321.4 and GMF 324.5-321.4b are 
not connected to any spiral arm.

\subsection{GMF 335.6-333.6}\label{sec:F333}

\begin{figure*}
\centering
\includegraphics[width = 0.9\textwidth]{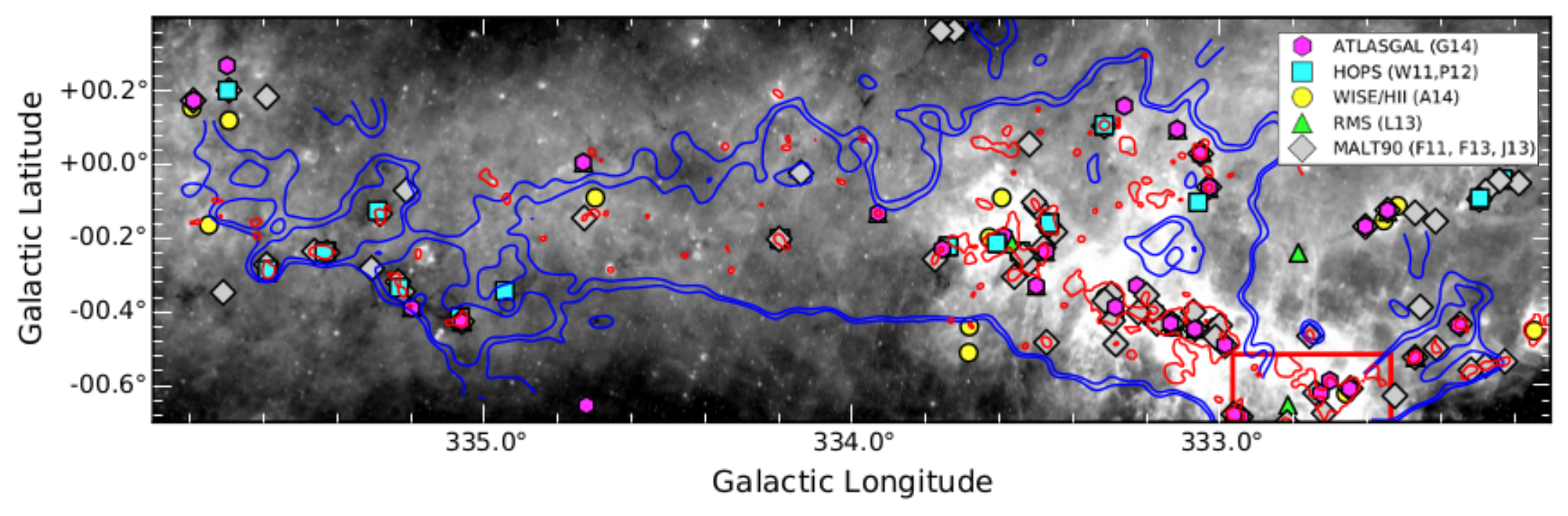}
\includegraphics[scale = 0.4]{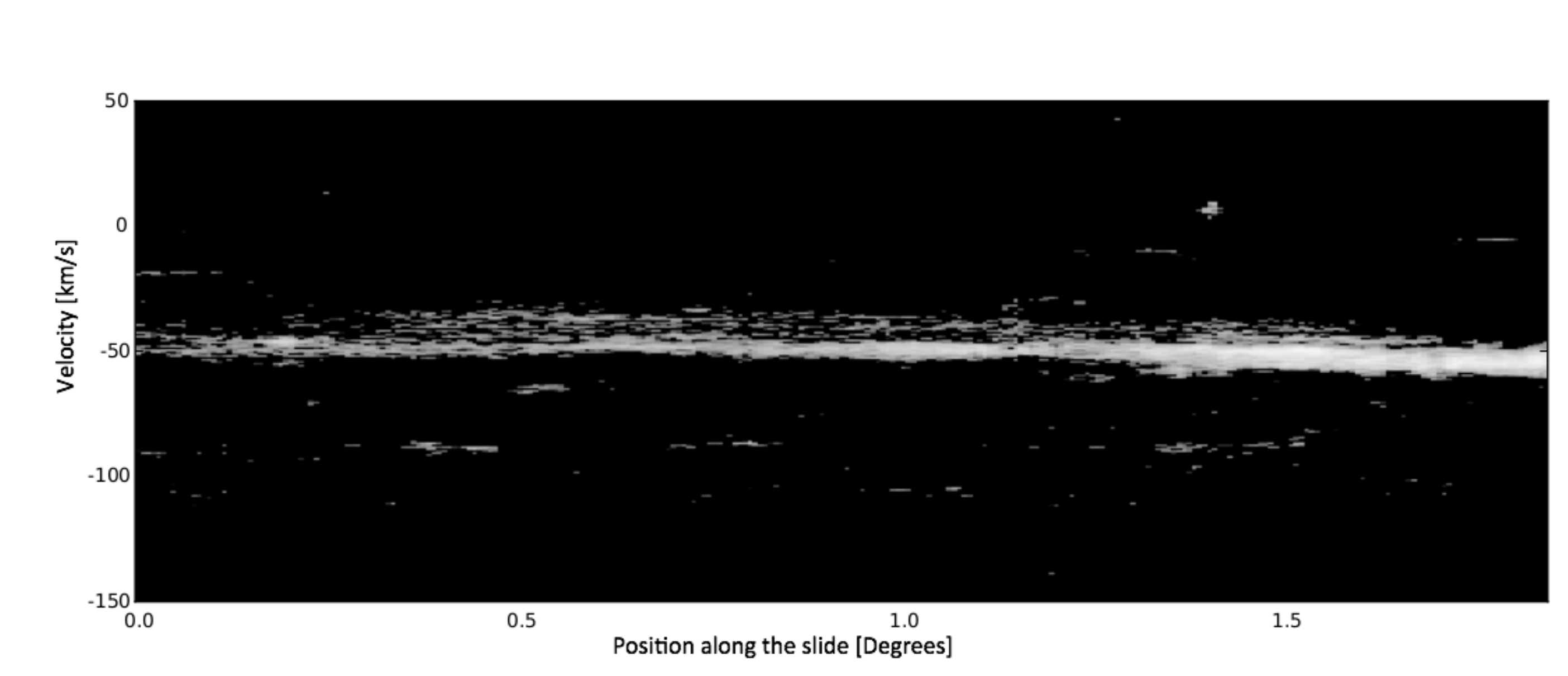}
\includegraphics[scale = 0.4]{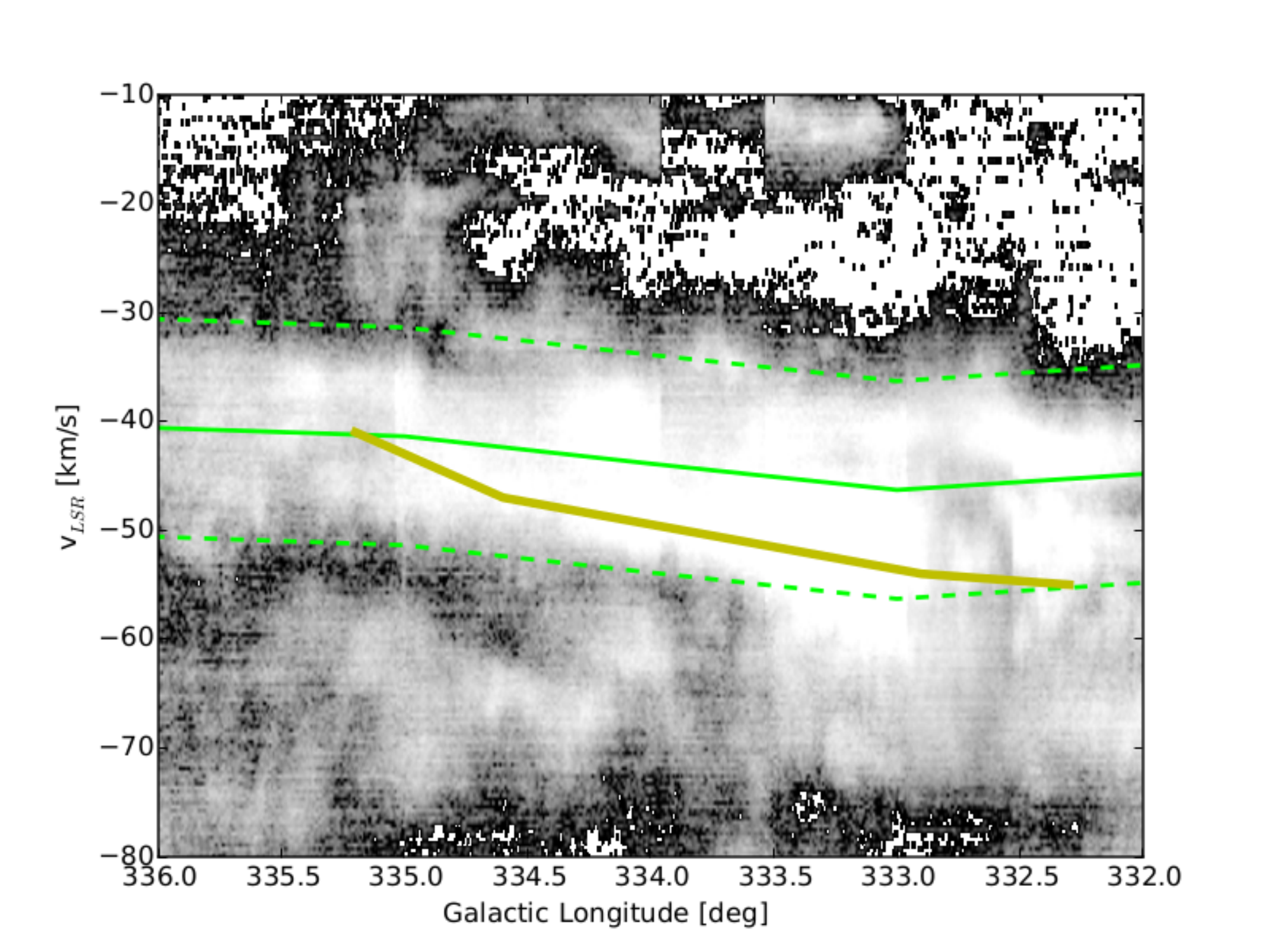}
\caption{\textit{Top:} Grayscale GLIMPSE $8\,\mu$m image of the GMF 335.6--333.6.
The blue contours show the \CO ~integrated intensity of 1.5\,K\,/km/s
and 2\,K\,/km/s,
integrated over the velocity range [-55,-35]\,km/s.
The red contours show the ATLASGAL emission at a contour level of 
$F_{870\mathrm{\,\mu m}}=250$\,mJy/beam. 
The red boxes show regions with poor ThruMMS data or absence of it. 
The filled geometric objects show all the dense gas measurements from 
different surveys with $v_{\mathrm{LSR}}$ within the velocity range  
indicated in Table~\ref{tab:filCandVelo}. Symbols as in Fig.~\ref{fig:apF3072_3054}.
\textit{Middle:} Position-velocity diagram of the \CO ~line of the GMF 335.6--333.6, obtained
 from a slice following the extinction feature used to identify 
 GMF 335.6--333.6. \textit{Bottom:}
PV diagram of the \tCO ~emission between $\lvert b \rvert \leq 1\degr$.
The yellow line shows GMF 335.6--333.6 in the PV space.
The green solid line shows the Scutum-Centaurus arm as predicted by~\citet{reid14}
and the dashed green lines show $\pm10$\,km/s of the velocity of the spiral arm..}
\label{fig:ap3356_3336}
\end{figure*}

It harbors one of the best studied \HII ~regions in the
southern Galactic hemisphere, G333 or RCW106~\citep{mookerjea04,roman09}. 
This \HII ~region is located at 3.6\,kpc (in agreement with the distance to
our GMF), it has a size of 30$\times$90\,pc and
a mass of $\sim2.7\times10^{5}_{\sun}$~\citep{bains06}.
GMF 335.6-333.6 is seen as an extinction feature
connected to RCW106.
The RCW106 complex is the main cause of the
remarkably high DGMF ($\sim37\%$) measured in GMF 335.6-333.6a.
The end at higher galactic longitudes is connected
to the S40 bubble. 
There is a small region of the GMF that is not covered by ThruMMS,
as it is shown with the red box in Fig.~\ref{fig:ap3356_3336}.
This GMF lies in the Scutum-Centaurus arm.

\subsection{GMF 335.6-333.6b}

This filament is very close to GMF 335.6-333.6a and it 
is recognizable as a prominent
extinction feature connecting two star-forming sites. It also
has velocity close to that of GMF 335.6-333.6.
Although in the first step we included GMF 335.6-333.6b as part of GMF 335.6-333.6,
we finally divided them for two reasons:
first, because they are not connected in \CO , and
second because their velocities slightly differ. GMF 335.6-333.6b
has also been identified as a Milky Way bone (see Sect.~\ref{sec:comp})
by~\citet{zucker15}. GMF 335.6-333.6b lies in the Scutum-Centaurus arm.

\begin{figure*}
\centering
\includegraphics[width = 0.9\textwidth]{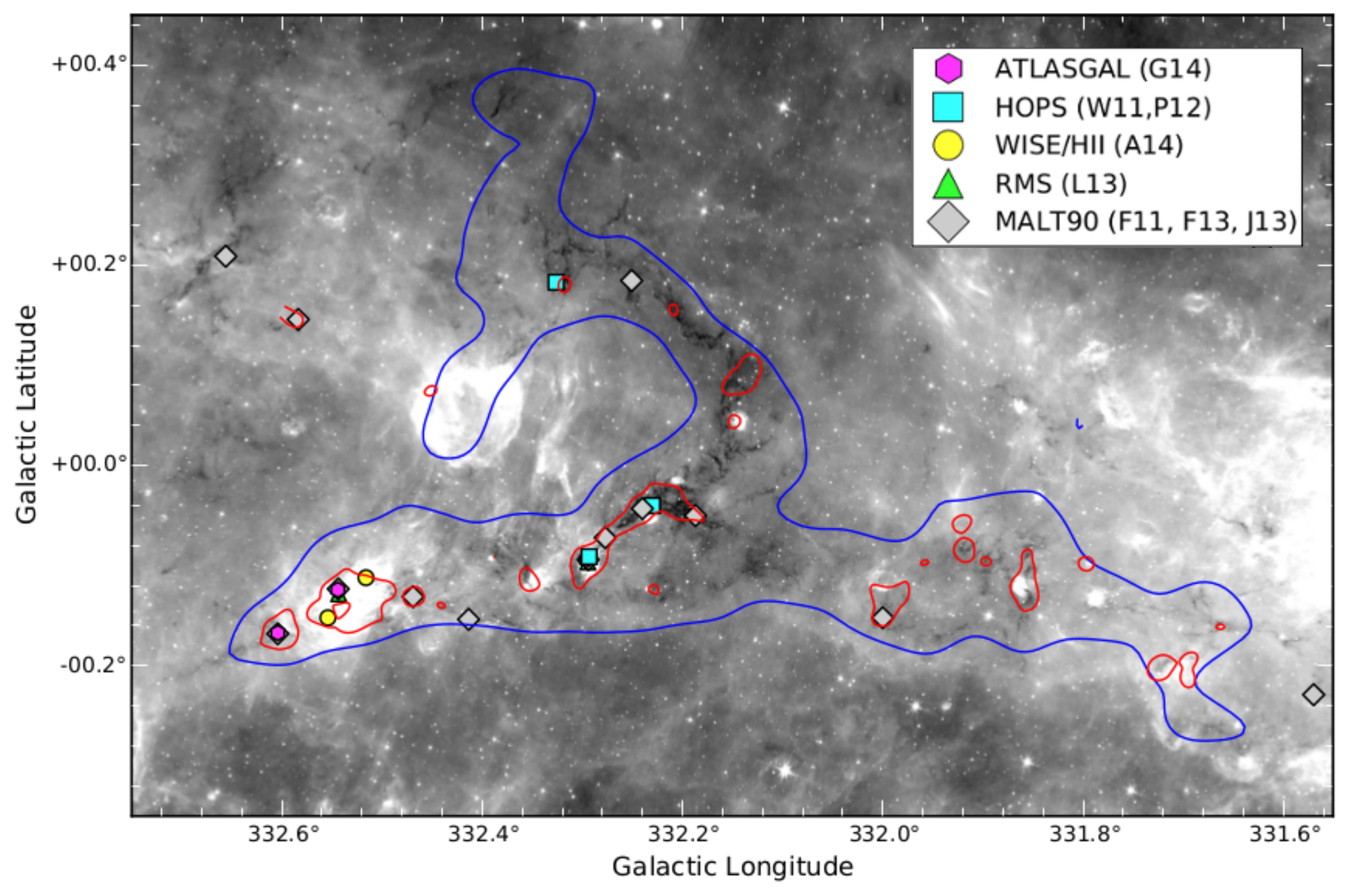}
\includegraphics[scale = 0.4]{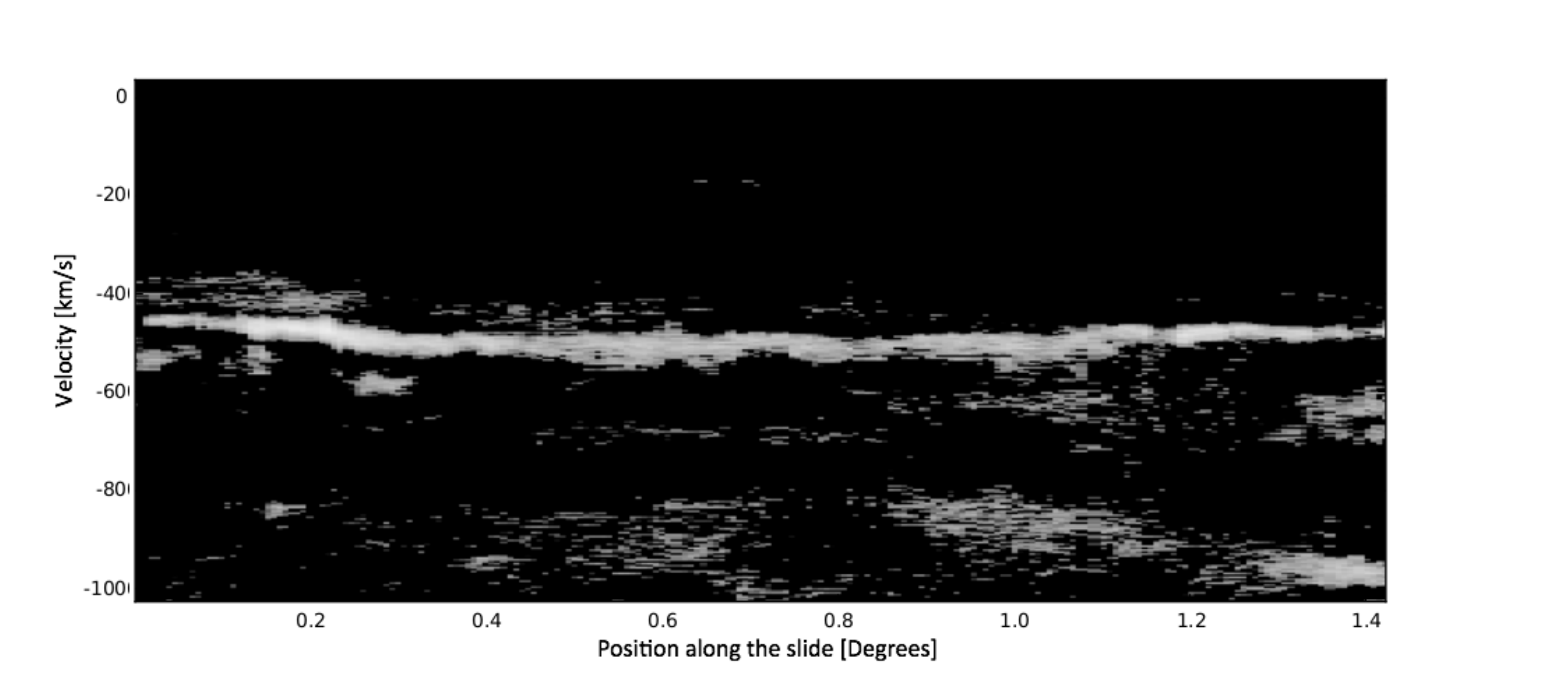}
\includegraphics[scale = 0.4]{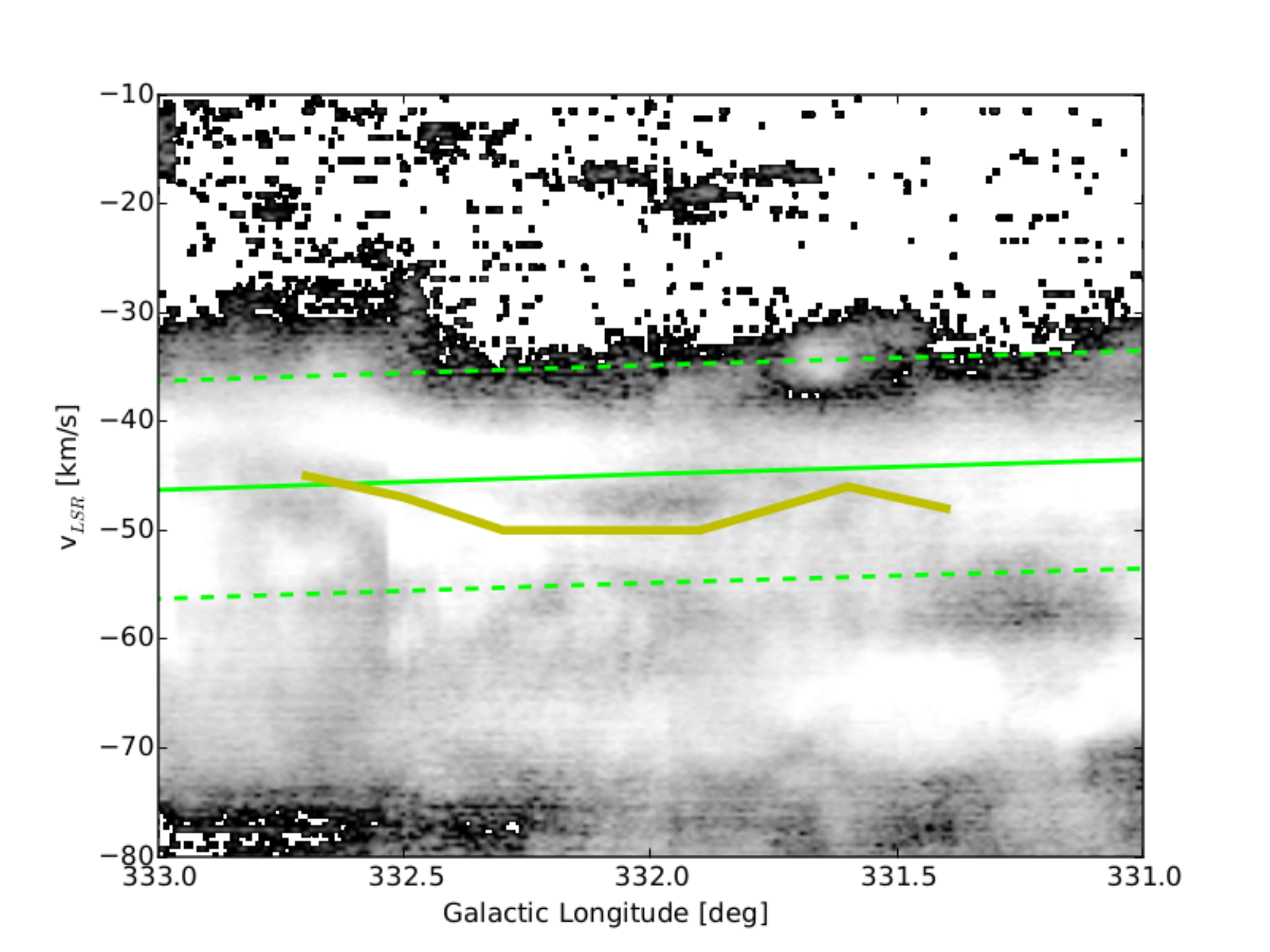}
\caption{\textit{Top:} Grayscale GLIMPSE $8\,\mu$m image of the GMF 335.6--333.6b.
The blue contours show the \CO ~integrated intensity of 3\,K\,/km/s,
integrated over the velocity range [-55,-45]\,km/s.
The red contours show the ATLASGAL emission at a contour level of 
$F_{870\mathrm{\,\mu m}}=250$\,mJy/beam. 
The filled geometric objects show all the dense gas measurements from 
different surveys with $v_{\mathrm{LSR}}$ within the velocity range  
indicated in Table~\ref{tab:filCandVelo}. Symbols as in Fig.~\ref{fig:apF3072_3054}.
\textit{Middle:} Position-velocity diagram of the \CO ~line of the GMF 335.6--333.6, obtained
 from a slice following the extinction feature used to identify 
 GMF 335.6--333.6b. \textit{Bottom:}
PV diagram of the \tCO ~emission between $\lvert b \rvert \leq 1\degr$.
The yellow line shows GMF 335.6--333.6b in the PV space.
The green solid line shows the Scutum-Centaurus arm as predicted by~\citet{reid14}
and the dashed green lines show $\pm10$\,km/s of the velocity of the spiral arm..}
\label{fig:ap3356_3336b}
\end{figure*}

\subsection{GMF 341.9-337.1}

\begin{figure*}
\centering
\includegraphics[width = 0.9\textwidth]{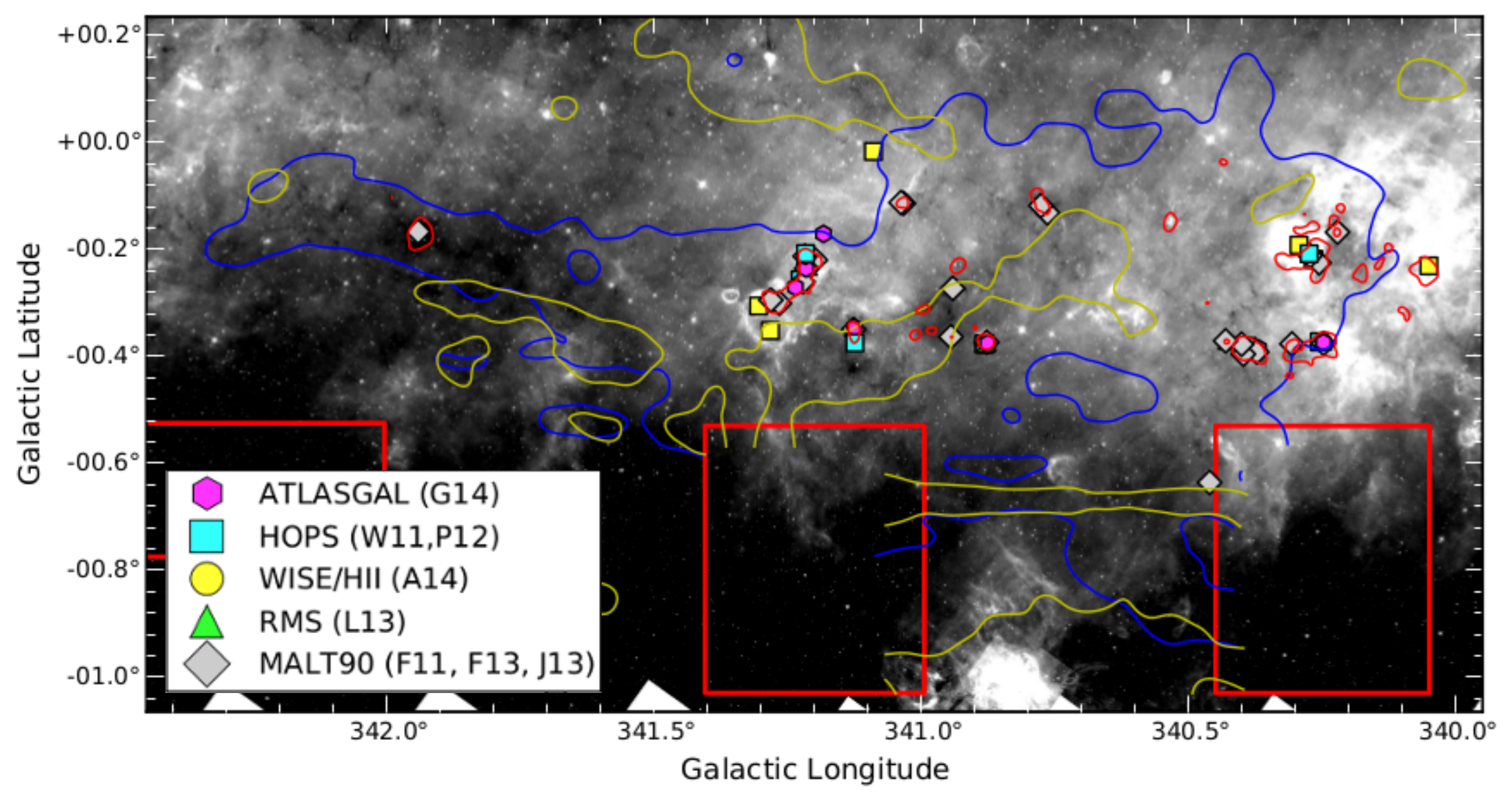}
\includegraphics[scale = 0.4]{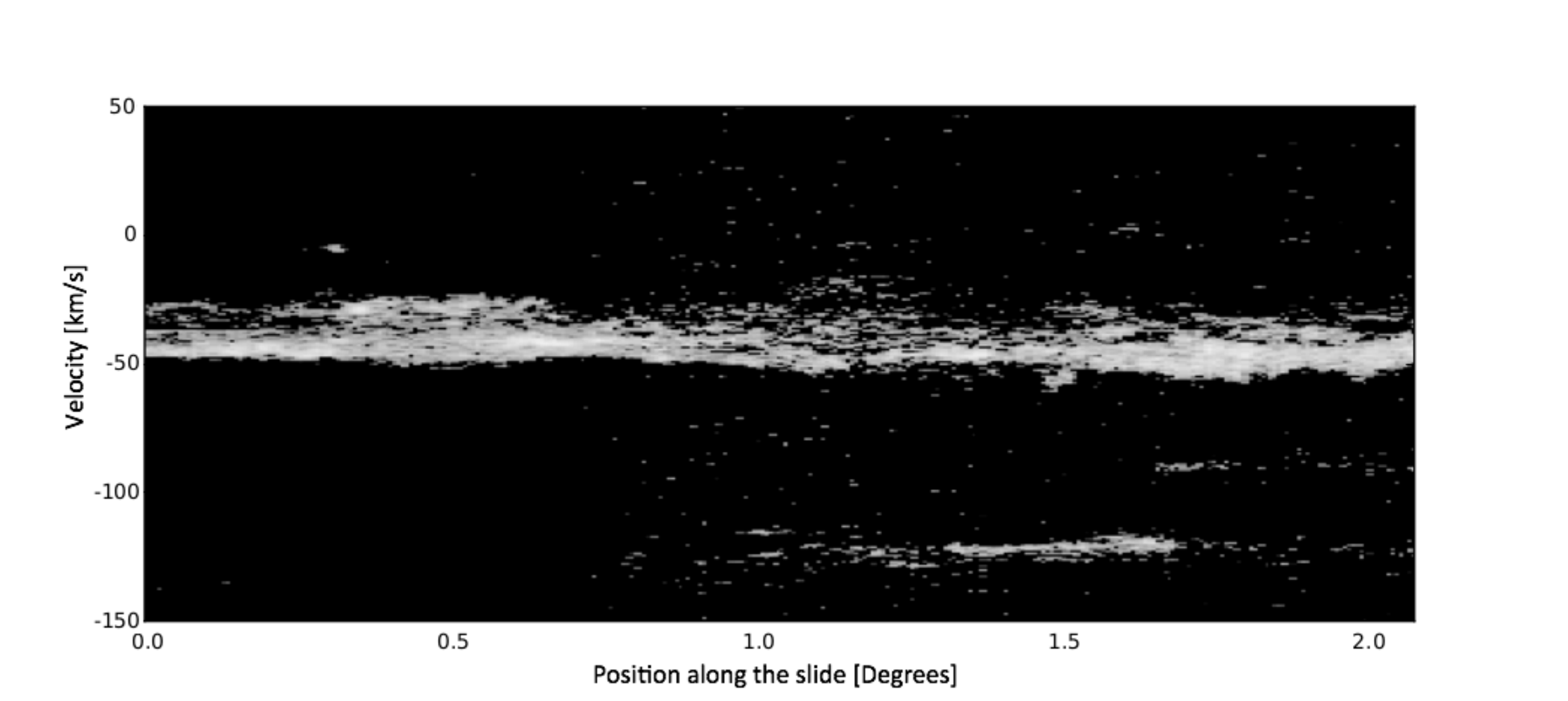}
\includegraphics[scale = 0.4]{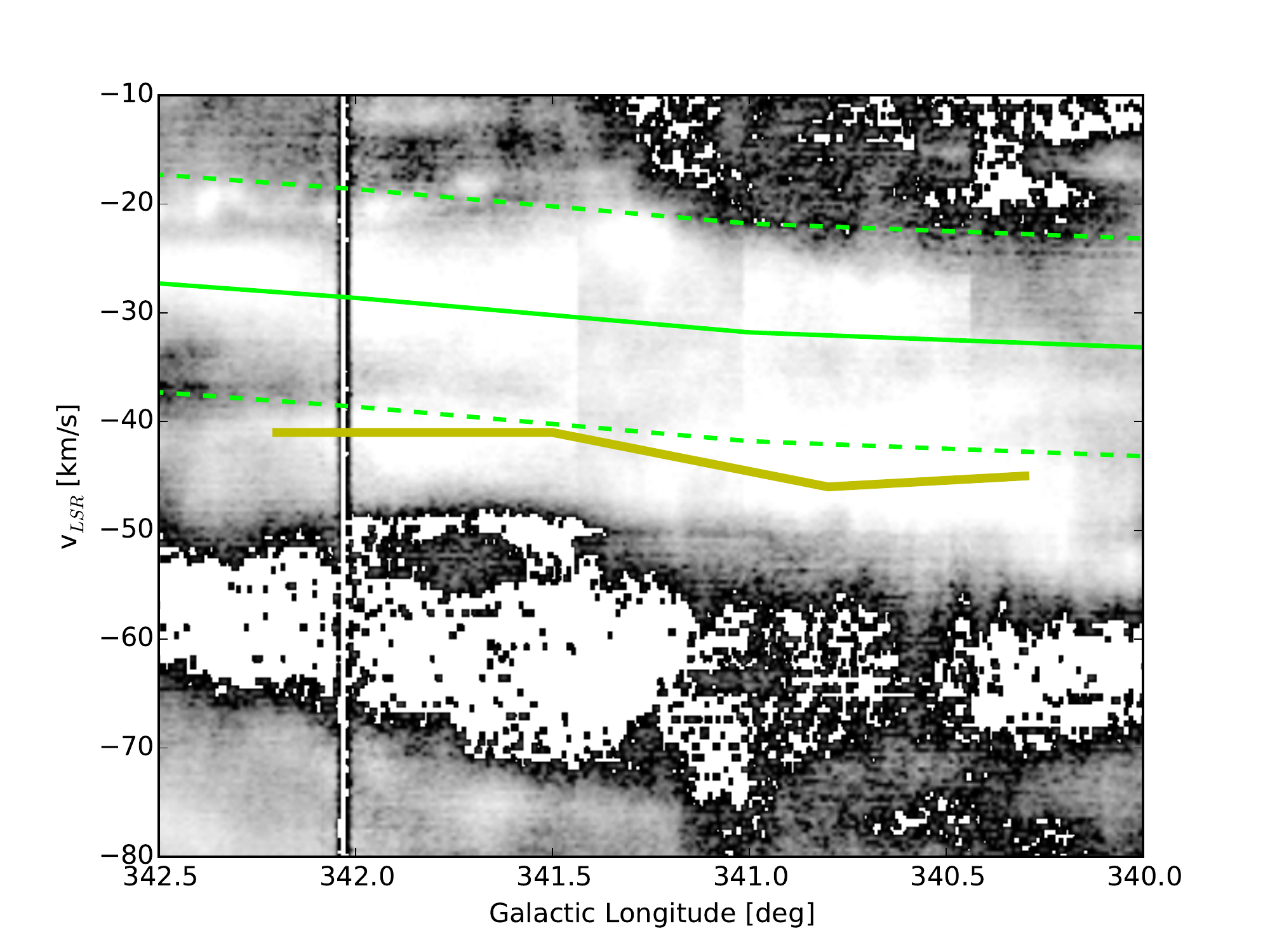}
\caption{\textit{Top:} Grayscale GLIMPSE $8\,\mu$m image of the GMF F341.9--337.1.
The blue and yellow contours show the \CO ~integrated intensity of 2\,K\,/km/s,
integrated over the velocity ranges [-47,-32]\,km/s and [-38,-20]\,km/s respectively.
The red contours show the ATLASGAL emission at a contour level of 
$F_{870\mathrm{\,\mu m}}=250$\,mJy/beam. 
The red boxes show regions with poor ThruMMS data or absence of it. 
The filled geometric objects show all the dense gas measurements from 
different surveys with $v_{\mathrm{LSR}}$ within the velocity range  
indicated in Table~\ref{tab:filCandVelo}. Symbols as in Fig.~\ref{fig:apF3072_3054}.
\textit{Middle:} Position-velocity diagram of the \CO ~line of the GMF 341.9--337.1, obtained
 from a slice following the extinction feature used to identify 
 GMF 341.9--337.1. \textit{Bottom:}
PV diagram of the \tCO ~emission between $\lvert b \rvert \leq 1\degr$.
The yellow line shows GMF 341.9--337.1 in the PV space.
The green solid line shows the Scutum-Centaurus arm as predicted by~\citet{reid14}
and the dashed green lines show $\pm10$\,km/s of the velocity of the spiral arm..}
\label{fig:ap3419_3371}
\end{figure*}

GMF 341.9-337.1 has a \HII ~region in its 
low-longitude end and another one, much more compact, in the middle. 
An extinction feature connects both star-forming sites and 
extends towards increasing longitudes. 
The southern part of the filament is not completely covered by ThruMMS
(see the red boxes in Fig.~\ref{fig:ap3419_3371}). As shown in
Fig.~\ref{fig:ap3419_3371}, almost every ATLASGAL clump in this filament has additional
kinematic information confirming the location of the dense gas 
inside GMF 341.9-337.1.

This filament seem to be part of Nessie extended
~\citep{goodman14}. However, GMF 341.9-337.1 does not lie in the 
Scutum-Centaurus arm, as Nessie extended
~does, but rather in an inter-arm region (see Fig.~\ref{fig:ap3419_3371}). 
We explore the reasons of this discrepancy below. Figure~\ref{fig:ap3419_3371}
shows that the emission of the Scutum-Centaurus arm 
has velocities of about [-35,-25]\,km/s, while the velocities
of GMF 341.9-337.1 are lower, [-48,-43]\,km/s.
This fact raises the question on whether GMF 341.9-337.1 is connected to
Nessie extended.

We integrated the \CO ~emission over the expected velocity range of 
Nessie extended to investigate 
whether it overlaps  with
GMF 341.9-337.1. In Fig.~\ref{fig:ap3419_3371} we show that 
both filaments overlap each other. However, there are extinction features 
that are only covered by GMF 341.9-337.1 and we also find that
most of the star-forming sites known are connected to GMF 341.9-337.1
rather than to Nessie extended.
We conclude that there are overlapping extinction features in this
region, better matched by GMF 341.9-337.1. This filament lies in an inter-arm region.

\subsection{GMF 343.2-341.7}

\begin{figure*}
\centering
\includegraphics[width = 0.9\textwidth]{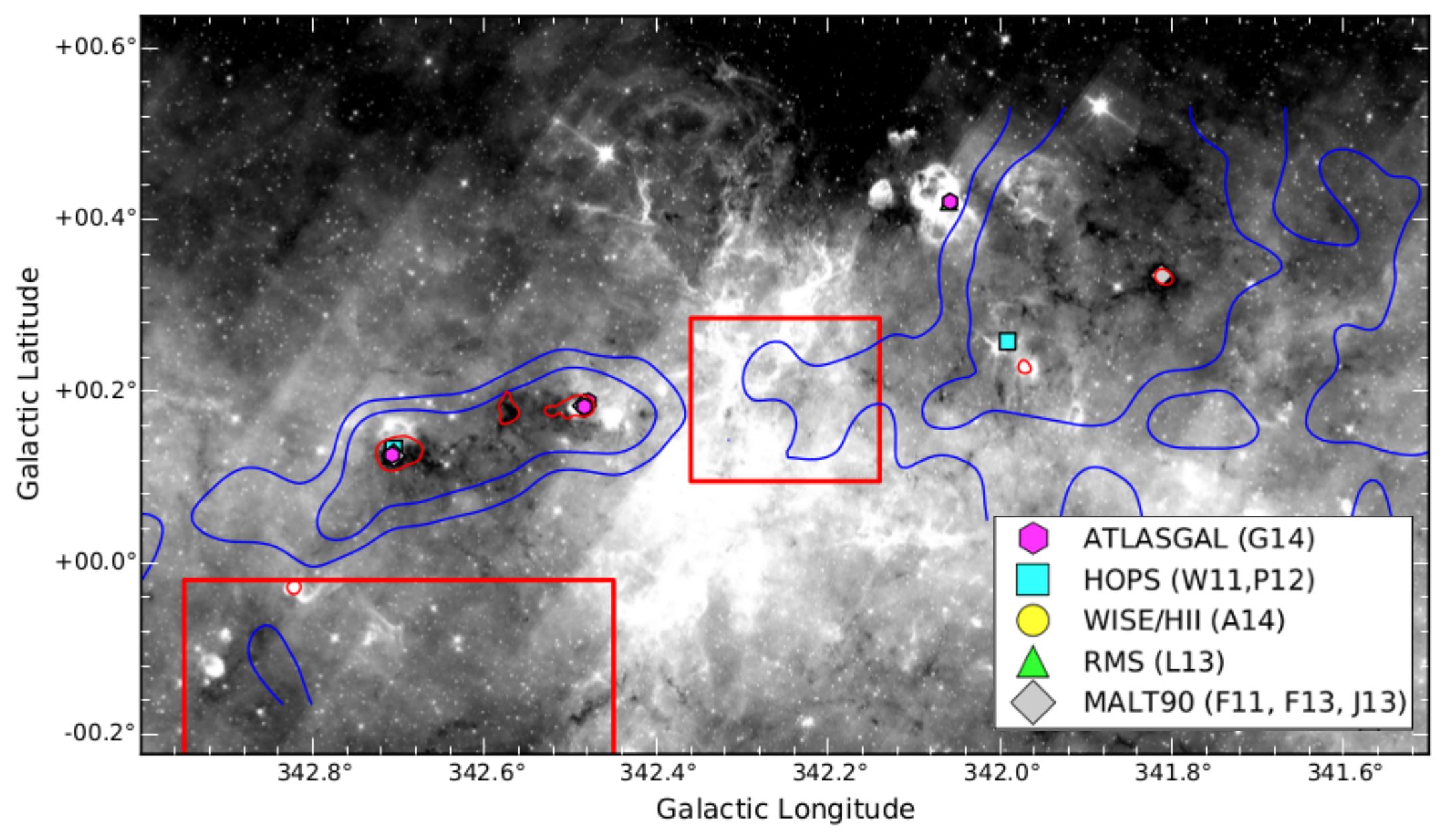}
\includegraphics[scale = 0.4]{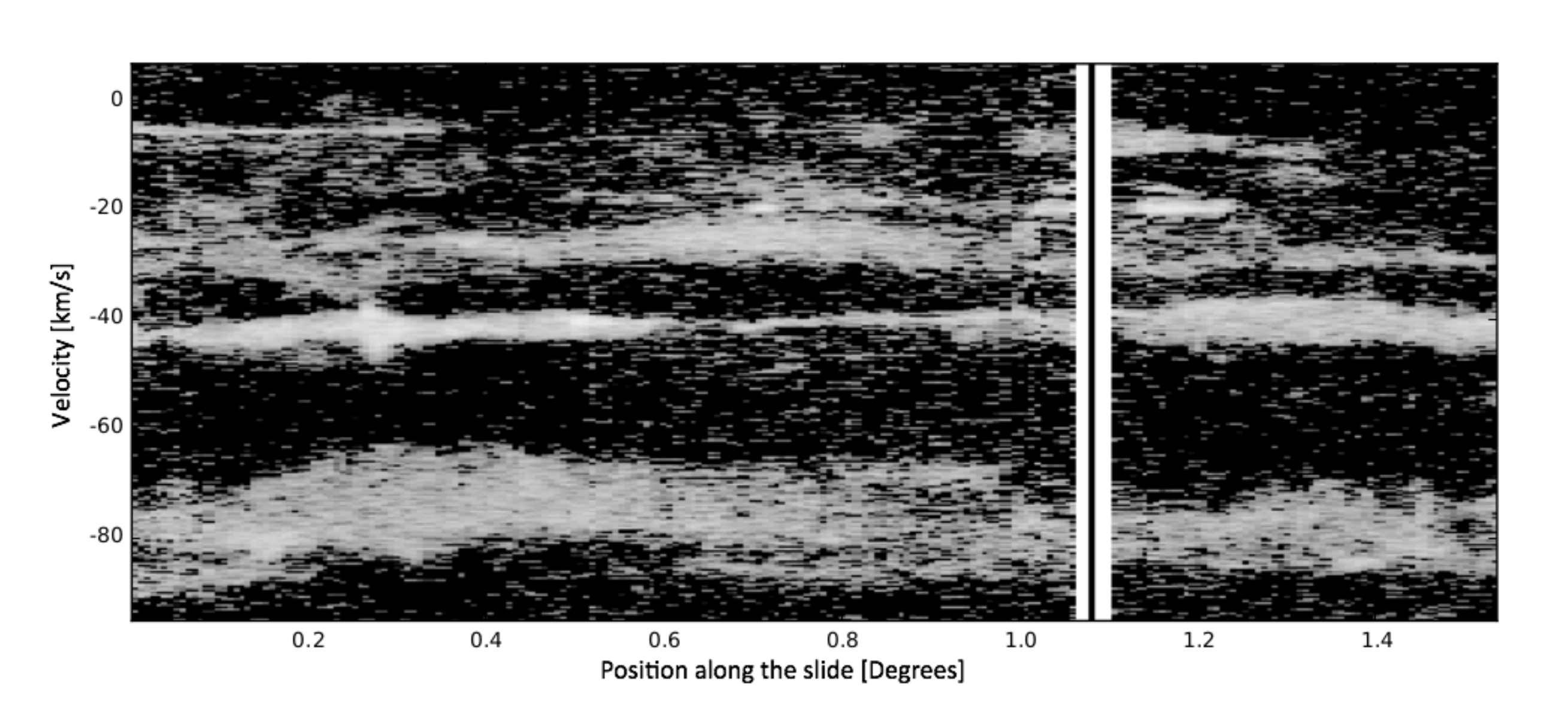}
\includegraphics[scale = 0.4]{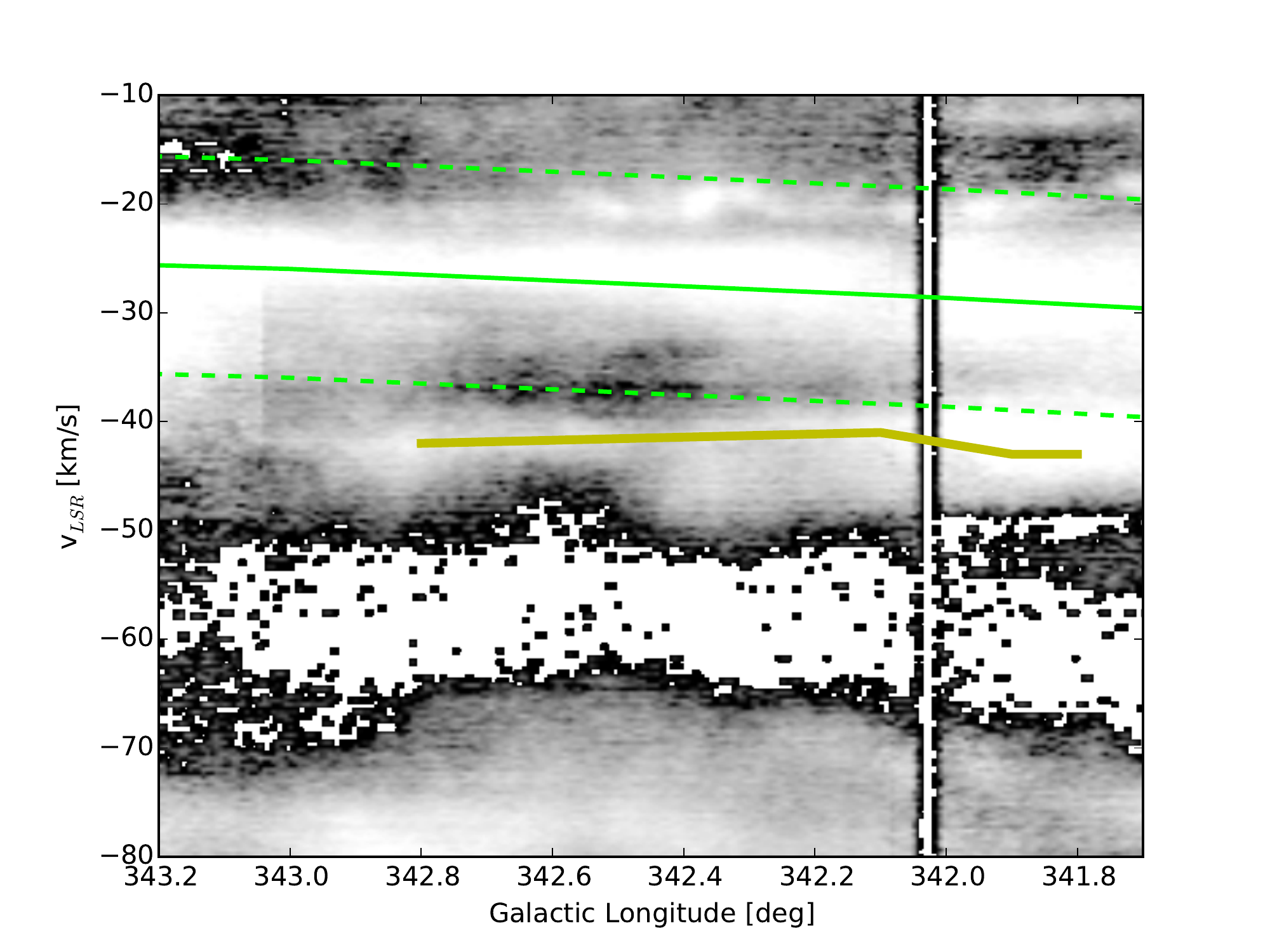}
\caption{\textit{Top:} Grayscale GLIMPSE $8\,\mu$m image of the GMF 343.2--341.7.
The blue contours show the \CO ~integrated intensity of 1.5\,K\,/km/s,
integrated over the velocity range [-50,-37]\,km/s.
The red contours show the ATLASGAL emission at a contour level of 
$F_{870\mathrm{\,\mu m}}=250$\,mJy/beam. 
The red boxes show regions with poor ThruMMS data or absence of it. 
The filled geometric objects show all the dense gas measurements from 
different surveys with $v_{\mathrm{LSR}}$ within the velocity range  
indicated in Table~\ref{tab:filCandVelo}. Symbols as in Fig.~\ref{fig:apF3072_3054}.
\textit{Middle:} Position-velocity diagram of the \CO ~line of the GMF 343.2--341.7, obtained
 from a slice following the extinction feature used to identify 
 GMF 343.2--341.7. \textit{Bottom:}
PV diagram of the \tCO ~emission between $\lvert b \rvert \leq 1\degr$.
The yellow line shows GMF 343.2--341.7 in the PV space.
The green solid line shows the Scutum-Centaurus arm as predicted by~\citet{reid14}
and the dashed green lines show $\pm10$\,km/s of the velocity of the spiral arm..}
\label{fig:apF3432_3417}
\end{figure*}

Two extinction features separated by a \HII ~region in 
projection. The high longitude end of the GMF is connected
to another smaller \HII ~region.
The significant \CO ~emission of this filament shows two
separated objects, although they are connected by significant \tCO ~emission. 
We note that there are high noise features in the ThruMMS data 
exactly in those \CO ~empty regions that
could hide a significant connection between the two parts of the \CO ~filament
(red boxes in Fig.~\ref{fig:apF3432_3417}). 
The high longitude section of the filament shows 
no dense gas, nor dense clumps connected to GMF 343.2-341.7. 
The GMF is not connected to any spiral arm, as it is shown
in the bottom panel of Fig.~\ref{fig:apF3432_3417}.

\subsection{GMF 358.9-357.4}

It lies in a very crowded region. Every spiral arm 
at these Galactic latitudes has velocities close to 0\,km/s.
This makes it difficult to isolate single velocity components.
The most prominent extinction feature in GMF 358.9-357.4 
is the IRDC known as G357~\citep{marshall09}. 
After isolating its velocity component
we found that it lies inside a much larger 
\CO ~complex. The distance estimates in this region are particularly 
hard to obtain using the~\citet{reid14} model. 
We therefore rejected this model for this filament
and instead used the distances from the literature: 3.3\,kpc~\citep{marshall09}.
This distance estimate places GMF 358.9-357.4 in the Scuttum-Crux arm.

\begin{figure*}
\centering
\includegraphics[width = 0.9\textwidth]{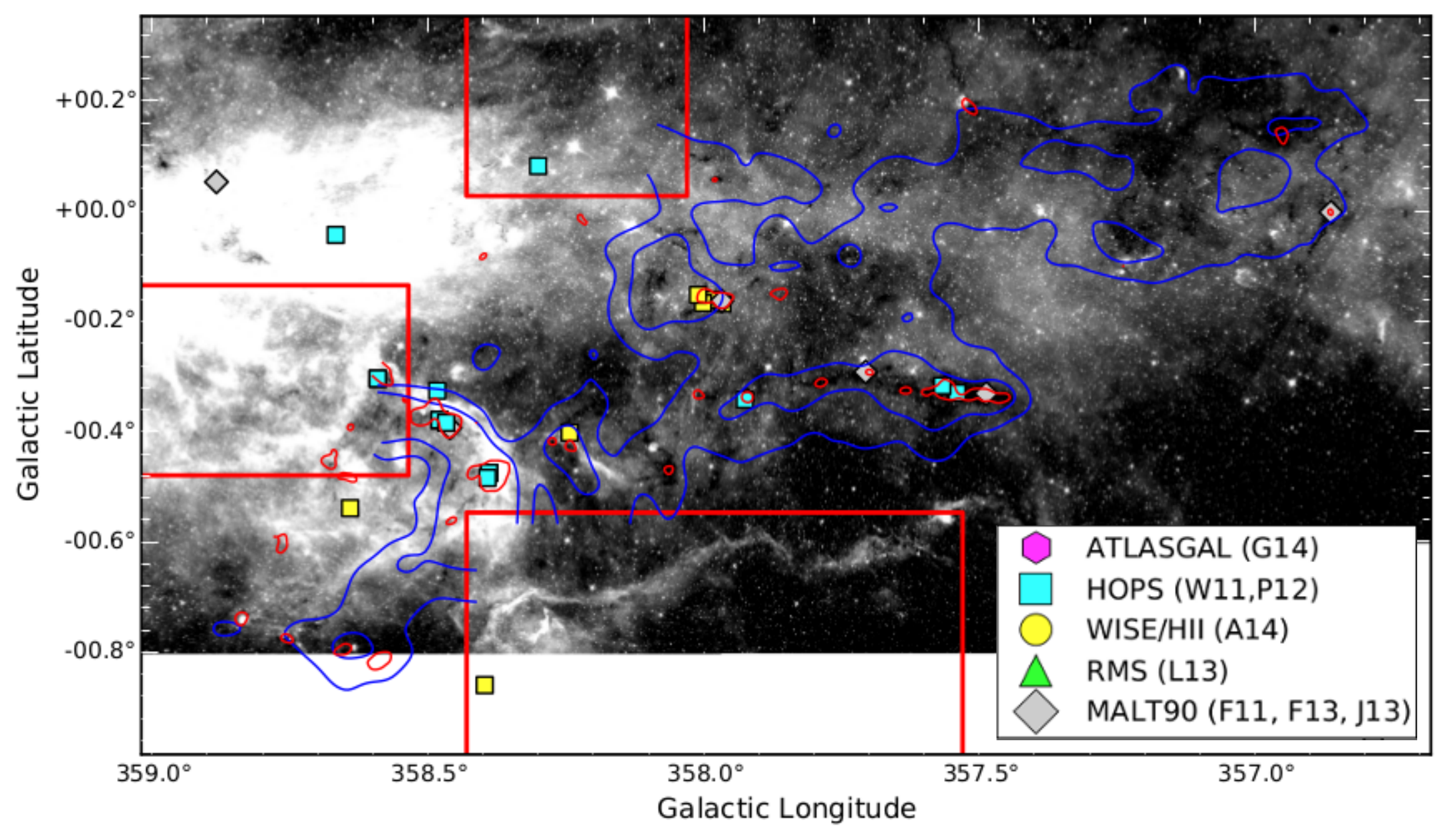}
\includegraphics[scale = 0.4]{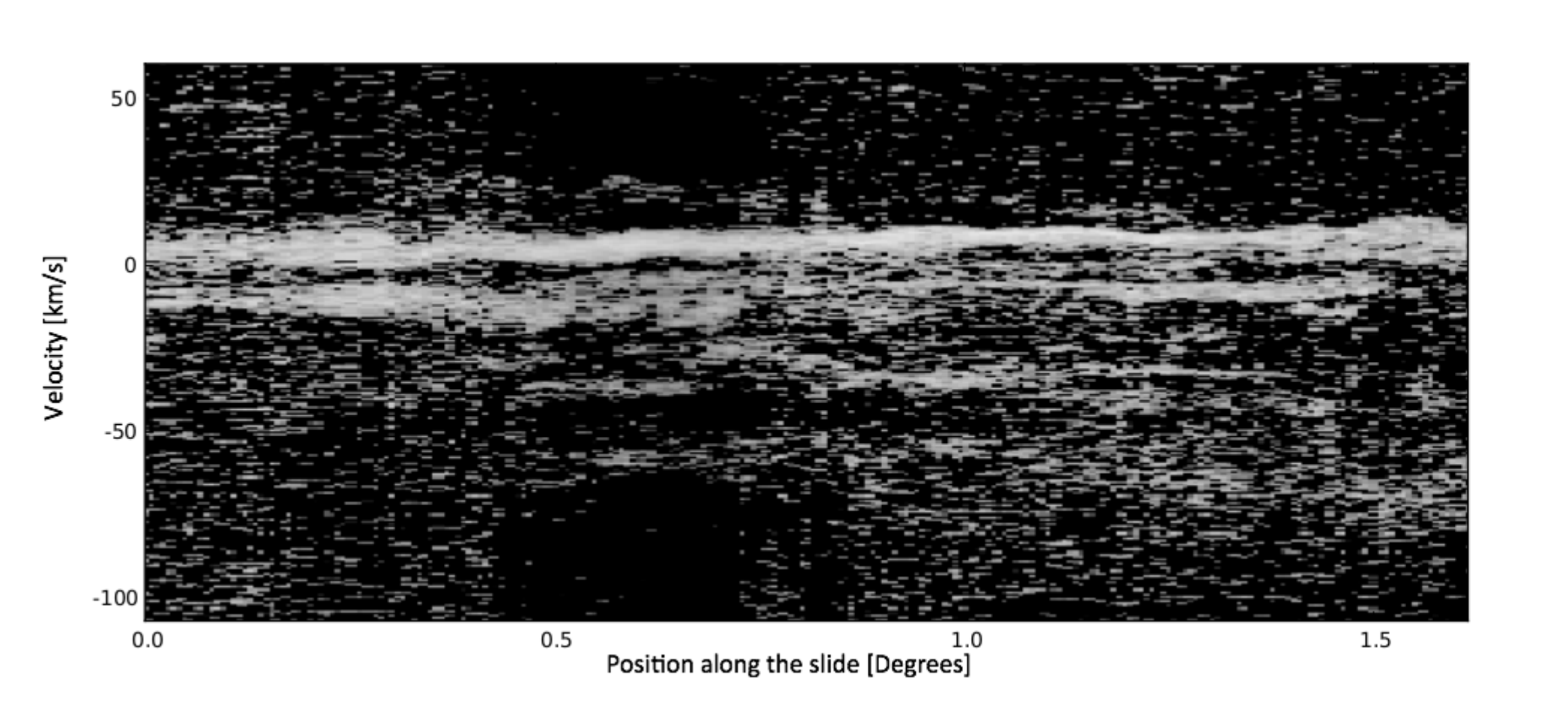}
\includegraphics[scale = 0.4]{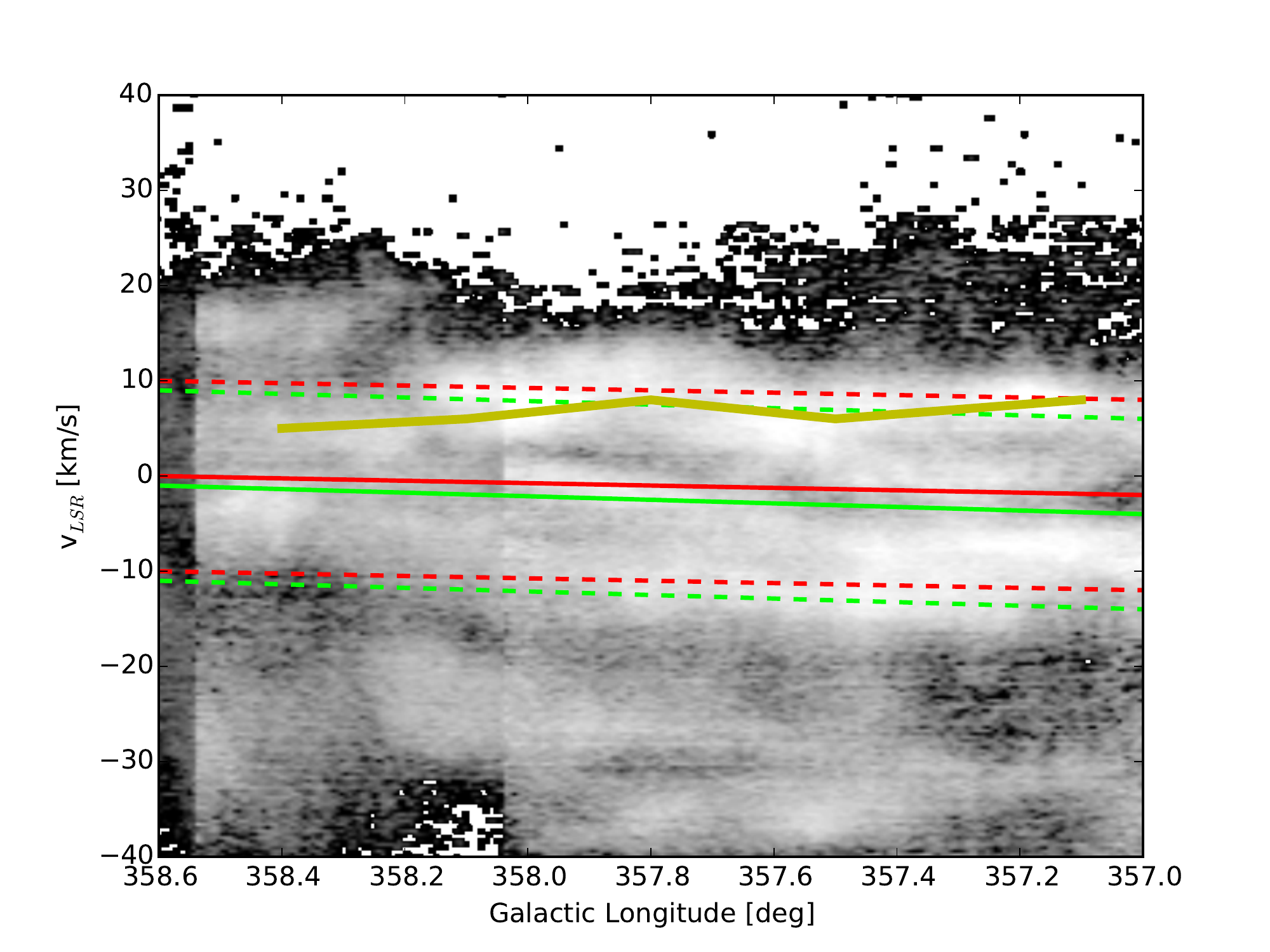}
\caption{\textit{Top:} Grayscale GLIMPSE $8\,\mu$m image of the GMF 358.9--357.4.
The blue contours show the \CO ~integrated intensity of 1.5\,K\,/km/s
and 3\,K\,/km/s,
integrated over the velocity range [0,10]\,km/s.
The red contours show the ATLASGAL emission at a contour level of 
$F_{870\mathrm{\,\mu m}}=250$\,mJy/beam. 
The red boxes show regions with poor ThruMMS data or absence of it. 
The filled geometric objects show all the dense gas measurements from 
different surveys with $v_{\mathrm{LSR}}$ within the velocity range  
indicated in Table~\ref{tab:filCandVelo}. Symbols as in Fig.~\ref{fig:apF3072_3054}.
\textit{Middle:} Position-velocity diagram of the \CO ~line of the GMF 358.9--357.4, obtained
 from a slice following the extinction feature used to identify 
 GMF 358.9--357.4. \textit{Bottom:}
PV diagram of the \tCO ~emission between $\lvert b \rvert \leq 1\degr$.
The yellow line shows GMF 358.9--357.4 in the PV space.
The green and red solid lines show respectively the Scutum-Centaurus and Sagittarius arms as predicted by~\citet{reid14}
and the dashed lines show $\pm10$\,km/s of the velocity of the spiral arms.}
\label{fig:apF3589_3574}
\end{figure*}

\onllongtab{
\begin{longtable}{ccccccc}
\caption{Dense gas tracers and star-forming signs associated to our GMFs}\\
\hline
\hline
$l$	&	$b$	&	$v_{\mathrm{LSR}}$	&	Filament assoc.	&	Object/line & Survey & Reference\\
$[\degr]$	&	$[\degr]$	&	$[km/s]$	&		&	& & \\
\hline
\endfirsthead
\caption{Continued.} \\
\hline
$l$	&	$b$	&	$v_{\mathrm{LSR}}$	&	Filament assoc.	&	Object/line & Survey & Reference\\
$[\degr]$	&	$[\degr]$	&	$[km/s]$	&		&	& & \\
\hline
\endhead
\hline
\endfoot
\hline
\endlastfoot
  305.18	&	    0.21	&	   -42.5	&	GMF\,3054\_ 3072	&	C$^{18}$O,$^{13}$CO,C$^{17}$O,$^{13}$C$^{18}$O & ATLASGAL & 1\\ 
  305.73	&	   -0.09	&	   -40.8	&	GMF\,3054\_ 3072	&	C$^{18}$O,$^{13}$CO,C$^{17}$O,$^{13}$C$^{18}$O & ATLASGAL & 1\\  
  305.17	&	   -0.00	&	   -34.2	&	GMF\,3054\_ 3072	&	C$^{18}$O,$^{13}$CO,C$^{17}$O,$^{13}$C$^{18}$O & ATLASGAL & 1\\  
  305.51	&	    0.02	&	   -39.8	&	GMF\,3054\_ 3072	&	C$^{18}$O,$^{13}$CO,C$^{17}$O,$^{13}$C$^{18}$O & ATLASGAL & 1\\  
  305.36	&	    0.20	&	   -37.5	&	GMF\,3054\_ 3072	&	NH$_{3}$ (1,1) and (2,2) & HOPS & 2 \\ 
  305.25	&	   -0.02	&	   -31.9	&	GMF\,3054\_ 3072	&	NH$_{3}$ (1,1) and (2,2) & HOPS & 2 \\ 
  305.22	&	    0.27	&	   -39.7	&	GMF\,3054\_ 3072	&	NH$_{3}$ (1,1) and (2,2) & HOPS & 2 \\ 
  304.55	&	    0.34	&	   -42.9	&	GMF\,3054\_ 3072	&	NH$_{3}$ (1,1) and (2,2) & HOPS & 2 \\
\end{longtable}
\tablebib{(1)~\citet{giannetti14};
(2)~\citet{purcell12}; (3)~\citet{anderson14};
(4)~\citet{lumsden13}; (5)~\citet{foster11,foster13,jackson13};
(6)~\citet{urquhart13a}; (7)~\citet{urquhart14b}; 
(8)~\citet{walsh11}; $^{(a)}$ See Table. 1 in~\citet{jackson13}.}
}

\end{document}